\documentclass[prb,twocolumn,groupaddress,showpacs] {revtex4}
\usepackage{amsmath,amsfonts,amssymb,mathbbol,dsfont}
\usepackage{graphicx,psfrag,color}
\usepackage{dcolumn}
\usepackage{bm}
\usepackage{mathbbol, appendix}
\usepackage{dsfont}

\def\dual{\ \stackrel{\Phi_\d}{\longrightarrow}\ }
\def\idual{\ \stackrel{\Phi^{-1}_\d}{\longrightarrow}\ }

\def\d{{{\sf d}}}

\def\r{{\bm{r}}}

\def\i{{\bm{e_1}}}
\def\j{{\bm{e_2}}}
 
\def\Z{{\mathbb{Z}}}

\def\Z{{\mathbb{Z}}}

\DeclareMathOperator{\e}{e}
\DeclareMathOperator{\de}{d\!}

\DeclareMathOperator{\sgn}{sgn}

\newcommand{\ket}[1]{|#1\rangle}

\begin{document}
\title{Topological phases in two-dimensional arrays of 
 parafermionic zero modes}
\author{M. Burrello}
\affiliation{Instituut-Lorentz, Universiteit Leiden, P.O. Box 9506, 2300 RA Leiden, 
The Netherlands.}
\author{B. van Heck}
\affiliation{Instituut-Lorentz, Universiteit Leiden, P.O. Box 9506, 2300 RA Leiden, 
The Netherlands.}
\author{E. Cobanera}
\affiliation{Instituut-Lorentz, Universiteit Leiden, P.O. Box 9506, 2300 RA Leiden, 
The Netherlands.}
 
\date{\today}

\begin{abstract}
It has recently been realized that zero modes with projective non-Abelian statistics, generalizing the notion of Majorana bound states, may exist at the interface between a superconductor and a ferromagnet along the edge of a fractional topological insulator (FTI). Here we study two-dimensional architectures of these non-Abelian zero modes, whose interactions are generated by the charging and Josephson energies of the superconductors. We derive low-energy Hamiltonians for two different arrays of FTIs on the plane, revealing an interesting interplay between the real-space geometry of the system and its topological properties. On the one hand, in a geometry where the length of the FTI edges is independent on the system size, the array has a topologically ordered phase, giving rise to a qudit toric code Hamiltonian in perturbation theory. On the other hand, in a geometry where the length of the edges scales with system size, we find an exact duality to an Abelian lattice gauge theory and no topological order.
\end{abstract}

\pacs{05.30.Pr, 74.81.Fa, 03.67.Lx, 11.15.Ha} 

\maketitle
 
\section{Introduction}

Systems that exhibit topological quantum order\cite{wen2004} have been a focus of attention in recent years. Part of the interest is due to the fact that they have been proposed as fault-tolerant quantum memories and platforms 
for quantum computation\cite{nayak2008}, the paradigmatic example being Kitaev's toric code\cite{kitaev2003}. The goal is to design architectures effectively governed by topologically-ordered Hamiltonians, where qubits may be stored and manipulated in a physically protected way.

Majorana zero modes, realized as superconducting midgap excitations in either one\cite{kitaev2001} or two \cite{volovik1999,read2000} spatial dimensions, are promising building blocks for such architectures. Two unpaired Majorana fermions at the ends of a one-dimensional (1D) superconducting wire encode non-locally a qubit\cite{kitaev2001} and, when allowed to move in a non-strictly 1D geometry, exhibit non-Abelian statistics\cite{moore1991,ivanov2001,alicea2011}, allowing to implement a non-universal set of quantum gates through ordered exchanges of their positions. Interest in Majorana fermions has increased considerably in recent times, since there are now several experimentally accessible systems that may host these quasiparticles (see Refs. \onlinecite{alicea2012} and \onlinecite{beenakker2012} for a review). A notable example is the edge of a two-dimensional (2D) topological insulator\cite{hasan2010,qi2011}, which hosts gapless helical (i.e., counter-propagating) modes, in proximity to an $s$-wave superconductor (SC) and a ferromagnet (FM). The competition between the proximity-induced SC and FM pairing along the edge results in the presence of a Majorana fermion at each domain wall\cite{fu2009}.

Recently, an interesting extension of this model was proposed in Refs. \onlinecite{stern2012, shtengel2012, cheng2012, vaezi2012}. While the edge excitations of a 2D topological insulator are normal electrons, it is possible to consider instead edge quasiparticles with a fractional charge $e/m$, where $m$ is an odd integer. Such gapless quasiparticles appear, for example, at the edge of the Laughlin fractional quantum Hall states, where they are described by a chiral Luttinger liquid theory\cite{wen1990,lee1991}. Due to the absence of time-reversal symmetry, these are chiral excitations. Helical $e/m$ quasiparticles would arise at the interface between two $\nu=1/m$ quantum Hall liquids with Land\'e $g$-factors of opposite sign or, similarly, as a Kramers doublet at the edge of a fractional topological insulator\cite{levin2009} (FTI). 

The simplest way to model FTIs is to consider them as fractional quantum spin Hall systems constituted by a two-dimensional gas of electrons subject to both a spin-dependent magnetic field (or a position-dependent spin-orbit coupling) and Coulomb interactions\cite{bernevig2006}. The first element creates two time-reversal symmetric Landau level structures, whereas the second gives rise to topologically ordered fractional states. These systems are gapped in the bulk (where Abelian anyonic excitations appear), but present fractional gapless edge modes.  While such time-reversal invariant topological phases have been thoroughly studied theoretically\cite{levin2009,levin2011,neupert2011,santos2011,levin2012}, no host material has emerged so far as an experimental candidate. We should also mention a recent proposal to realize a fractional helical liquid in quantum wires\cite{oreg2013}.

Along the FTI edges, the proximity effect with superconductors and ferromagnets results in the presence of zero modes\cite{stern2012,shtengel2012}. Since the second-quantization operators associated with these zero modes inherit a fractional exchange phase $(2\pi)/(2m)$ from the unperturbed edge fields, we shall refer to them as $\Z_{2m}$ parafermionic (PF) zero modes. They are projective non-Abelian anyons\cite{barkeshli2012,you2012}, with fusion rules that generalize those of Majorana fermions, affording extended computational power\cite{stern2012,shtengel2012,barkeshli2012,hastings2012,you2013}.

These superconducting zero modes realize a 1D model with $\mathbb{Z}_{2m}$ symmetry studied 
by Fendley \cite{fendley2012}, which is an extension of Kitaev's Majorana 
chain model\cite{kitaev2001} and hosts PF zero modes 
localized at the edges of the system. While the Kitaev chain is dual to the quantum Ising 
chain via a Jordan-Wigner transformation, the $\mathbb{Z}_{2m}$ chain model is dual to the 
1D chiral Potts ($p$-clock) model, with $p=2m$. 

Indeed, PFs as collective degrees of freedom are indeed well-known in statistical mechanics\cite{mussardo}. They appear naturally in the study of the 2D $p$-state clock models \cite{fradkin1980,rajabpour2007,ortiz2012} and their quantum 1D counterparts\cite{cobanera2011}. In lattice systems, they arise as products of order and disorder operators defined for self-dual systems. Moreover, PFs admit a description in terms of a $\mathbb{Z}_p$ invariant conformal field theory\cite{zamolodchikov1985} (CFT)  featuring the PFs as primary fields. PF zero modes in superconducting systems are however related only to CFTs with unit central charge (see, for example, Ref. \onlinecite{delfino2012}), arising from a bosonization description of the FTIs edge modes.

In light of this body of research, it is interesting to extend the recent works on $\Z_{2m}$ PFs to 2D networks of superconductors. For Majorana zero modes, this question has already been addressed in literature\cite{xu2010,terhal2012,nussinov2012}. Majorana lattice models can be mapped into Ising models, allowing for a description of their phase diagram. They can exhibit topologically ordered phases and realize the toric code in a perturbative regime\cite{terhal2012}. The extension of these analyses to $\Z_{2m}$ PFs may reveal new Hamiltonian realizations of fault-tolerant stabilizer codes\cite{bravyi1998,bravyi2010} for quantum bits with \(2m\) states, and hence novel platforms for quantum memories generalizing the toric code \cite{bullock2007}. However, for $\Z_{2m}$ PFs the extension to 2D lattices is less immediate than in the Majorana case, partially because of the connection to clock models, which are less well understood than Ising models.

\begin{figure}[t!]
\includegraphics[width=\columnwidth]{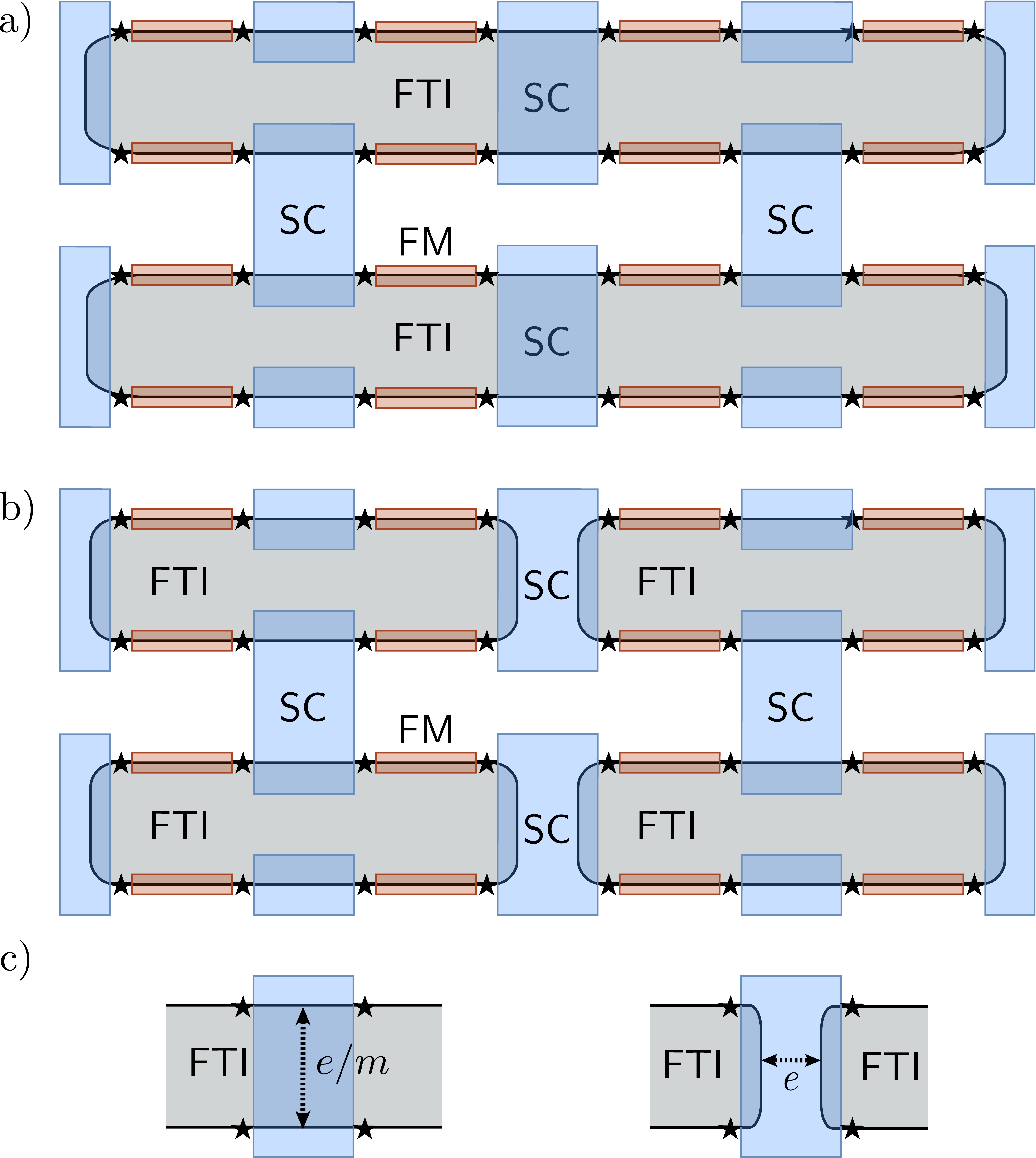}
\caption{The two different 2D architectures considered in this paper. They are composed of superconductors (SC, blue) and ferromagnets (FM, red) deposited on top of a 2D array of fractional topological insulators (FTI, grey). $\Z_{2m}$ PF zero modes, marked as black stars, arise at each SC/FM interface along the edge. We consider two possible geometries: in panel (a), the FTIs extend for the whole length of the system, while in panel (b) the FTIs have fixed size. If we enlarged on the horizontal direction the system in panel (a), the number of FTIs would stay constant and the edge length would increase, while the vice versa would happen in panel (b). In the main text, we refer to the two architectures as the {\it stripe} and {\it tile} models respectively. As shown schematically in panel (c), the two models can also be distinguished by different tunneling regimes between the two edge segments gapped out by the same superconducting island. If the SC covers a single FTI (left), tunneling of fractional charge $e/m$ may take place between the two edges, while if the SC covers two different FTIs (right), only electron tunneling is allowed, since transport of a fractional charge cannot take place via a topologically trivial bulk. The tile model (b) only has SC islands of this second kind, while the stripe architecture (a) has both.}
\label{fig_layout}
\end{figure}

In order to fill this gap, in this work we consider two distinct 2D architectures of $\Z_{2m}$ PFs, shown in Fig.~ \ref{fig_layout}. The architectures are obtained from a pattern of superconductors and ferromagnets layered on top of an array of 2D FTIs. The only difference between the two models is the geometry of the underlying FTI array. In Fig.~ 1(a) the array is formed by long stripes of FTIs extending for the whole length of the system, while in Fig.~ 1(b) the stripes are cut in smaller pieces (or tiles) of fixed dimension. For ease of discussion, we shall refer to these two architectures as the \textit{stripe} and \textit{tile} model respectively. 

Similarly to Refs. \cite{xu2010, terhal2012, nussinov2012}, the effective Hamiltonian of the two models is dictated by two mesoscopic phenomena:
\begin{enumerate}
	\item the fractional Josephson effect, mediated by the tunneling of $e/m$ quasiparticle between two different superconductors, and
	\item the charging energy of the superconductors.
\end{enumerate}
The fractional Josephson effect arising with $\Z_{2m}$ PF zero modes has already been investigated in Refs.\cite{shtengel2012, cheng2012}, while to our knowledge the interplay between PF zero modes and Coulomb energy was not considered in previous works.

While both the stripe and the tile architectures give rise to a square lattice of $\Z_{2m}$ PF zero modes, and despite the fact that the effective Hamiltonian contains the same set of local interactions in both models, we will show that the different geometry of the FTI edges is crucial to determine their properties, which turn out to be quite distinct. Indeed, since different edge geometries yield a different set of commutation rules for the $\Z_{2m}$ PF operators and different physical constraints on the Hilbert space, they can determine different topological properties.

The paper is organized as follows: in Sec.~\ref{sec_2Dmodel} we derive the effective
Hamiltonian for the stripe and tile architectures, considering both Josephson and Coulomb energies, and explain the physical constraints and conservation laws specific to 
each array. In Sec.~\ref{secJW} we map the effective Hamiltonian into two different 
clock models, using two non-isomorphic sets of PF Jordan-Wigner transformations. We analyze 
the phase diagram of the two models in Sec.~\ref{secphase}, where we show that the 
tile model realizes a qudit toric code Hamiltonian in perturbation
theory while the stripe model is dual to a gauge theory. We conclude with an outlook 
in Sec.~\ref{sec_conclusion}.

\section{Effective Hamiltonian for 2D parafermionic architectures}\label{sec_2Dmodel}

\begin{figure*}[t!]
\centering
\includegraphics[width=\textwidth]{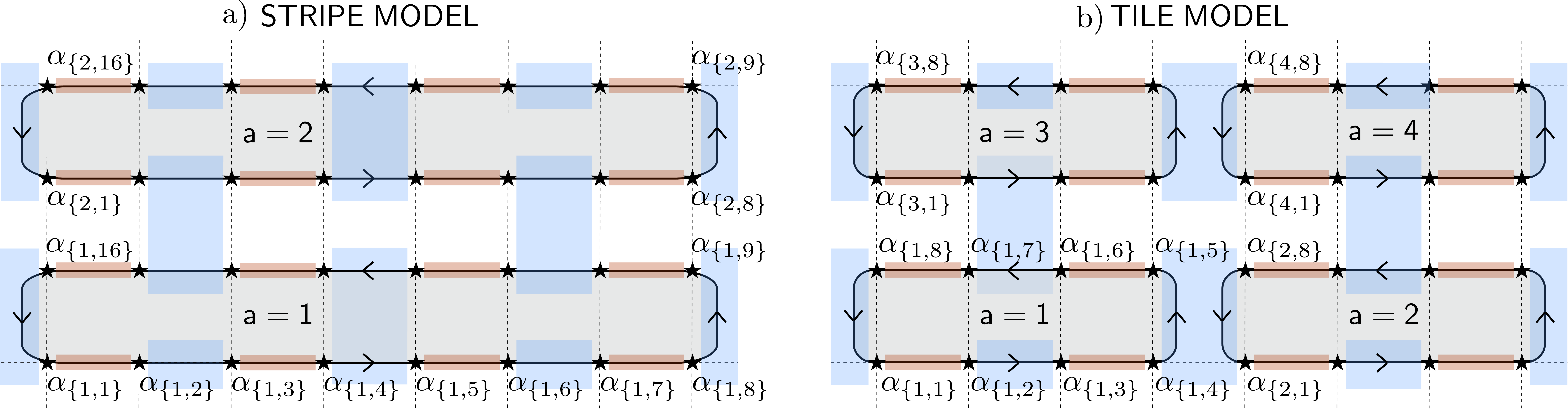}
\caption{A square lattice of $\Z_{2m}$ PF zero modes of dimensions $L_x=8$ and $L_y=4$. To label the PFs, we follow the convention established in the main text: first we order the FTI edges with an index ${\sf a}$, and then we order the PFs along each edge with an index $j$, starting from an arbitrary origin. Each PF zero mode is then denoted as $\alpha_{\{{\sf a},j\}}$, and all commutation rules between operators at different sites are fixed unambiguously.}
\label{fig_notation}
\end{figure*}

In the two architectures in Fig.~ \ref{fig_layout}, each FM/SC interface along the edge of a FTI hosts a PF zero mode. Hence the stripe and tile architectures generate arrays of interacting $\Z_{2m}$ PFs, which are protected by the superconducting and ferromagnetic gaps and thus determine the low-energy physics of the system. In this section we derive an effective Hamiltonian in terms of $\Z_{2m}$ PF operators for a square lattice of dimensions $L_x\times L_y$ with open boundaries\cite{LxLy}. Generalizations to other boundary conditions can be easily implemented. 

Each PF zero mode is described by a second-quantization operator $\alpha$ satisfying the relations:
\begin{align}
&\alpha^{2m}=1, \label{prop1}\\
&\alpha^\dagger=\alpha^{2m-1}.\label{prop2}
\end{align}
We can associate to $\alpha$ and $\alpha^\dagger$ respectively the annihilation and creation of a charge $e/m$ on the adjacent superconductor, in such a way that a Cooper pair is split in $2m$ quasiparticles\cite{stern2012}.  Eqs. (\ref{prop1}-\ref{prop2}) can be derived from the Luttinger liquid description of the FTI fractional edges, as done in detail in Refs.~\onlinecite{stern2012}, \onlinecite{shtengel2012} and as outlined in Appendix \ref{bosonization_appendix}.

Furthermore, these PF operators obey 
unconventional commutation rules. Denoting two different PF operators with 
generic, ordered labels $\mu$ and $\nu$, we have
\begin{align}
\alpha_\mu\alpha_\nu&=\e^{-i\epsilon_{\mu\nu}\pi/m}\,
\alpha_\nu\alpha_\mu,\label{commrules}\\
\alpha^\dagger_\mu\alpha_\nu&=\e^{+i\epsilon_{\mu\nu}\pi/m}\,
\alpha_\nu\alpha_\mu^\dagger ,\label{commrules2}
\end{align}
where $\epsilon_{\mu\nu}=-\epsilon_{\nu\mu}=\pm1$ is a sign that must be fixed by convention.
As we outline in Appendix \ref{bosonization_appendix}, Eqs. (\ref{prop1}-\ref{commrules2}) 
can be derived from the underlying helical Luttinger liquid theory for the FTI edges. 
Note that for $m=1$ the $\epsilon_{\mu\nu}$'s do not play any role and the equations (\ref{prop1}-\ref{commrules2})  reproduce all the properties of Majorana fermions. 

In the 1D case, $\mu$ and $\nu$ are integers denoting the positions of the PFs on a line. All signs are fixed by assigning an orientation to the line, so that $\epsilon_{\mu\nu}=\sgn(\mu-\nu)$. In two dimensions the ordering procedure is slightly more complicated and proceeds in the following way.
\begin{enumerate}
	\item We label each FTI edge of the system with an integer ${\sf a}$, thus introducing an ordering of the edges. We also assign a counterclockwise orientation to each edge ${\sf a}$.
	\item Starting from an arbitrary origin and following the counterclockwise orientation, we label all ferromagnets along the edge with an integer $k=1,\dots,M$ (similarly to what was done in Ref. \onlinecite{stern2012}). The number $M$ is the total number of FMs along each FTI edge: note that $M=4$ for the tile model, while $M=L_x$ for the stripe model (see Fig.~ \ref{fig_notation}).
	\item We identify the SC/FM interfaces at the left and the right of each FM with an integer $j=2k-1$ or $j=2k$ respectively.
\end{enumerate}
The PFs $\alpha_\mu, \alpha_\nu$ are thus labelled by a composite index $\mu=\{{\sf a}, j\}$, $\nu=\{{\sf a}', j'\}$ and we fix all the conventional signs as
\begin{align}\label{signs}
\epsilon_{\mu\nu}=\sgn({\sf a}-{\sf a}')+\delta_{{\sf a}{\sf a}'}\sgn(j-j').
\end{align}
In Fig.~ \ref{fig_notation} we explicitly illustrate the procedure for labeling all the PFs of our square array, in both the tile and stripe architecture. Due to the different number and disposition of the FTIs, the PFs in the stripe model are actually distinct from (non-isomorphic to) the PFs in the tile model. The value of $\epsilon_{\mu\nu}$ may differ for pairs of PFs in the same site of the square lattice, and consequently the set of commutation relations Eqs. (\ref{commrules},\ref{commrules2}) is not the same for the two geometries. From the point of view of the physical components, this difference can be traced back to the following fact [see also Fig.~ \ref{fig_layout}(c)]. In the tile model, the array is fully constituted by SC islands connecting two different FTIs. Quasiparticle tunneling from one FTI edge to the other is forbidden since the two edges are separated by a topologically trivial region. The stripe model, instead, is composed also by a second type of SC island, connecting two edges of the same FTI. In this case, tunneling of a charge $e/m$ from one edge segment to the other is possible, albeit suppressed by the bulk gap, akin to what happens in a fractional quantum Hall constriction \cite{chamon1993}.

In order to describe physical interactions between PFs, it is useful to introduce the operator
\begin{equation} \label{zcharge}
P_{\mu\nu}=\e^{i\epsilon_{\mu\nu}\pi/2m}\alpha^\dagger_\nu\alpha_\mu,
\end{equation}
defined for every given pair $\alpha_\mu, \alpha_\nu$.  For $m=1$, $P_{\mu\nu}$ represents 
the $\Z_2$ fermionic parity associated with two Majorana fermions. Here we are extending 
this notion to the $\mathbb{Z}_{2m}$ symmetry of the PFs, and we shall refer to 
$P_{\mu\nu}$ as $\mathbb{Z}_{2m}$ charge. From its Hermitian conjugate,
\begin{equation}
P_{\mu\nu}^\dagger=\e^{-i\epsilon_{\mu\nu}\pi/2m}\alpha^\dagger_\mu\alpha_\nu,
\end{equation}
we see that it is a unitary operator,
\begin{equation}
P_{\mu\nu}P^\dagger_{\mu\nu}=P^{\dagger}_{\mu\nu} P_{\mu\nu}=1.
\end{equation}
Moreover $P_{\mu\nu}^{2m}=1$. Thus its eigenvalues must be the $(2m)$-th roots of unity,
\begin{equation}\label{eigen_phases}
\lambda_n=\e^{in\pi/m},\quad n=0,1,\cdots, 2m-1.
\end{equation}
A pair $\alpha_\mu$, $\alpha_\nu$ can be irreducibly represented on a $2m$-dimensional Hilbert space, with a basis given by 
the states $\ket{n}$ such that $P_{\mu\nu}\ket{n}=\lambda_n\ket{n}$. The Hilbert 
space dimension of a square lattice of PF zero modes of size $L_x\times L_y$ is
therefore $(2m)^{L_x\cdot L_y/2}$.

Now that we have set the basic algebraic rules, we can write down the physical ingredients 
of the model - Josephson and charging energy. These will form the basic local bonds used 
to write an effective 2D Hamiltonian for the PFs. 

\subsection{Josephson energy}

Thanks to the presence of zero modes, phase-coherent tunneling of $e/m$ quasiparticles may 
take place across the ferromagnetic region between adjacent superconductors along a 
common edge. The resulting Josephson effect is characterized by an anomalous periodicity 
of $4\pi m$ 
(in units of the superconducting flux quantum $\Phi_0=h/2e$), essentially because the tunneling quasiparticle's charge is reduced by a factor $2m$ with respect to the charge of a Cooper 
pair\cite{fu2009,stern2012,shtengel2012,cheng2012}. In other words, the anomalous period reflects the fact that the junction can be in $2m$ different states associated to the $\Z_{2m}$ charge of two PFs located at its ends. Physically, these states are distinguished by the fractional spin of the ferromagnet inside the junction\cite{stern2012}, or equivalently by the number of the fractional quasiparticles trapped in it (modulo $2m$).

Using the notation introduced in Fig.~ \ref{fig_notation}, the $\mathbb{Z}_{2m}$ 
charge of a junction situated on edge ${\sf a}$ can be written as
\begin{equation}\label{junction_parity}
P_{\left\lbrace {\sf a},2k-1\right\rbrace, 
\left\lbrace {\sf a},2k\right\rbrace }=\e^{-i\pi/2m}\,
\alpha^\dagger_{\left\lbrace {\sf a},{2k}\right\rbrace }
\alpha_{\left\lbrace {\sf a},{2k-1}\right\rbrace }.
\end{equation}
It acts as a transfer operator, destroying one $e/m$ charge inside the superconductor on one side of the junction and creating it on the other side. Such tunneling processes can be modeled by an effective Hamiltonian of the form
\begin{equation}\label{JJham}
H_{{\sf J}}=-\frac{J}{2}\left(\e^{i(\delta-\pi)/2m}\,
\alpha^\dagger_{\left\lbrace {\sf a},{2k}\right\rbrace }
\alpha_{\left\lbrace {\sf a},{2k-1}\right\rbrace }+{\rm H.c.}\right).
\end{equation}
Here $J$ is the tunneling strength and $\delta$ is the phase difference between the two superconductors. The tunneling Hamiltonian splits the states of the junction in $2m$ energy branches given by
\begin{equation}\label{JJspectrum}
E_{{\sf J},n}=-J\,\cos\,\left(\frac{\delta}{2m}+\frac{n\pi}{m}\right)
\end{equation}
with $n=0, \dots, 2m-1$.

As in the case of Majorana zero modes \cite{fu2009,xu2010}, the fractional Josephson effect mediated by PF modes prevails over the ordinary Josephson effect mediated by Cooper pairs, which is a higher-order effect. Moreover, the addition of the ordinary Josephson term, with $2\pi$ periodicity in the phase difference, would not modify qualitatively our results, thus it is neglected here and in the following. 

\subsection{Charging energy}

Let us now consider a single superconducting island of our array, and let us denote with 
$\phi$ and $N=-2i\,\tfrac{\de}{\de\phi}$ the phase and number operators of this island.
The presence of the PF zero modes becomes manifest through non-trivial (twisted) 
boundary conditions in the condensate ground state wave-function $\Psi(\phi)$\cite{fu2010},
\begin{equation}\label{twisted_bc}
\Psi(\phi+2\pi)=e^{i\pi q}\,\Psi(\phi)\ .
\end{equation}
Here $q$ represents the charge in units of $e$ inside the superconductor (modulo $2e$). 
The spectrum of the number operator depends on these twisted boundary conditions, since its twisted eigenfunctions 
\(\chi_s(\phi)=e^{i(s+q/2)\phi}/\sqrt{2\pi}\) satisfy
\begin{equation}
N\chi_s=(2s+q)\chi_s,\quad s\in \Z.
\end{equation}
In a conventional superconductor, $q=0$, we have periodic 
boundary conditions, and \(N\) counts Cooper pairs. In the presence of Majoranas $q$ may assume either 
value $\{0,1\}$ giving periodic or anti-periodic boundary conditions depending on the 
fermion parity of the superconductor\cite{fu2010}. In the presence of $\Z_{2m}$ PF zero modes, the possible values of 
$q$ are extended to fractional values:
\begin{equation}
q=\{\tfrac{n}{m}\}=\{0, \tfrac{1}{m},\tfrac{2}{m}, \dots, 1, \tfrac{m+1}{m},\dots,
\tfrac{2m-1}{m}\}.
\end{equation}
The resulting boundary conditions are twisted with possible phases 
$\e^{in\pi/m}$, and the spectrum of $N$ is given by 
rational numbers with denominator $m$. 

Ground states with different values of $q$ are not anymore degenerate if the charging energy of the superconducting island,
\begin{equation}
H_{{\sf ch}}=E_C\,\left(N-n_{{\sf ind}} \right)^2\,,
\end{equation}
is taken into account. Here $E_C=e^2/2C$, $C$ is the self-capacitance of the superconductor,
and $n_{\sf ind}$ the charge (in units of $e$) induced on the island by nearby voltage gates.

For our purposes, it is useful to separate the contribution of the fractional charges to the charging energy from that of the Cooper pairs. We will therefore work in a regime which highlights the role of the former, as done in Ref. \onlinecite{vanheck2012} for Majorana fermions. If all superconducting islands are connected to a grounded superconductor via a conventional Josephson junction of energy $E_J\gg E_C$, the superconducting phases are pinned to their classical minima, freezing the bosonic degree of freedom associated with Cooper pairs. The charging energy splits the ground state degeneracy by inducing quantum phase slips. In this semiclassical regime, $H_{\sf ch}$ can be 
replaced by an effective Hamiltonian of the form\cite{koch2007}
\begin{equation}\label{HDelta}
H_{\Delta}=-\Delta \cos(\pi q+\pi n_{\sf ind})\ .
\end{equation}
The cosine dependence on the charge in this effective Hamiltonian is reminiscent 
of the Aharonov-Casher effect\cite{aharonov1984}. When a (Josephson) vortex encircles 
a superconducting island, it acquires a phase proportional to the charge contained in it. 
The energy $\Delta$ is exponentially small in the ratio $E_J/E_C$.

Let us now write explicitly the interaction \eqref{HDelta} in terms of the PF operators. We denote as $q_{{\sf a},k}$ the fractional charge trapped inside the segment of an FTI edge ${\sf a}$ between the $k$-th and $(k+1)$-th ferromagnet. As such, it can be expressed as
\begin{equation} \label{zchargea}
\e^{i\pi q_{\mathsf{a},k}}\equiv P_{\left\lbrace \mathsf{a},2k \right\rbrace \left\lbrace 
\mathsf{a},2k+1 \right\rbrace } = \e^{-i\frac{\pi}{2m}} \alpha_{\left\lbrace 
\mathsf{a},2k+1\right\rbrace }^\dag \alpha_{\left\lbrace \mathsf{a},2k\right\rbrace } .
\end{equation}
In the special case $k=M$, Eq. \eqref{zchargea} has to be supplemented with the boundary condition
\begin{equation} \label{alpha_boundary}
\alpha_{\left\lbrace \mathsf{a},2M+1\right\rbrace } = 
\e^{-i\pi q_{\sf a}} \alpha_{\left\lbrace \mathsf{a},1\right\rbrace },
\end{equation}
where $q_{\sf a}$ is the total fractional charge along the edge ${\sf a}$. Eq. \eqref{alpha_boundary} appears naturally in the bosonization description of the PFs as the boundary condition of a closed edge with total charge $q_{\sf a}$ surrounding no net magnetic flux,\cite{stern2012} see also Appendix \ref{bosonization_appendix}. This boundary condition constitutes a constraint that the physical states of the system must fulfill, as we will discuss more extensively in the final part of this section. 

In our 2D architecture, each SC gaps out either one or two segments of an FTI edge, depending on whether it lies at the boundary of the system or in the bulk. In the second case, the total fractional charge $q$ contained in it is the sum of two charges $q_{{\sf a}, k}$, $q_{{\sf a}', k'}$ and can be expressed as $\e^{i\pi q}=\e^{i\pi (q_{{\sf a}, k}+q_{{\sf a}', k'})}=P_{\left\lbrace \mathsf{a},k \right\rbrace \left\lbrace \mathsf{a},k+1 \right\rbrace }P_{\left\lbrace \mathsf{a}',k' \right\rbrace \left\lbrace \mathsf{a}',k'+1 \right\rbrace }$, since two $\Z_{2m}$ charges operators always commute if they do not share a PF operator. The charging energy takes the form
\begin{widetext}
\begin{equation}\label{HD}
H_\Delta=\begin{cases}-\dfrac{\Delta}{2}\,\left(\e^{i\pi n_{\sf ind}}\e^{-i\pi/2m}\,\alpha_{\left\lbrace 
\mathsf{a},2k+1\right\rbrace }^\dag \alpha_{\left\lbrace \mathsf{a},2k\right\rbrace } +\rm{h.c.}\right)&\qquad \text{on the boundary} \\\\-\dfrac{\Delta}{2}\,\left(\e^{i\pi n_{\sf ind}}\e^{-i\pi/m}\,\alpha_{\left\lbrace 
\mathsf{a},2k+1\right\rbrace }^\dag \alpha_{\left\lbrace \mathsf{a},2k\right\rbrace }\,\alpha_{\left\lbrace 
\mathsf{a}',2k'+1\right\rbrace }^\dag \alpha_{\left\lbrace \mathsf{a}',2k'\right\rbrace }+\rm{h.c.}\right)&\qquad\text{in the bulk}\end{cases}
\end{equation}
\end{widetext}
The total charges $q_{\sf a}$ of the FTI edges may appear in the Hamiltonian \eqref{HD} as additional phases, due to Eq. \eqref{alpha_boundary}.

\subsection{Effective Hamiltonian}

Adding together the contributions from all islands and junctions, we arrive to an effective Hamiltonian
\begin{equation}\label{Heff}
H=\sum_{{\sf islands}} H_{\Delta} + \sum_{{\sf junctions}} H_{{\sf J}}\;.
\end{equation}
Each PF of the array belongs to one superconducting island and one junction and therefore it appears twice in the effective Hamiltonian, once in $H_{\Delta}$ and once in $H_{\sf J}$. 

Note that the effective Hamiltonian is the same for the stripe and the tile 
architectures, which share the same lattice, the same number of parafermions 
and the same set of local interactions. Nevertheless, the presence of two different sets 
of commutation rules for the PF operators is enough to give the two systems markedly 
different properties. 

\subsection{Conserved quantities and charge constraints} \label{sec_constraints}

The two different commutation rules between PFs in the stripe and tile architectures
are due to the fact that the Hilbert spaces of the two system are constrained
in physically different ways. To see this, notice that the total charge $q_{\sf a}$ at 
the edge of each FTI $\sf{a}$ must be conserved since 
no term in the Hamiltonian \eqref{Heff} introduces tunneling between different 
fractional topological insulators. That is,
\begin{equation}\label{conservation_law}
[\e^{i\pi q_{\sf a}},\,H]=0
\end{equation}
for every $\sf{a}$. Moreover, the total charge of each FTI (edge plus bulk) is not only conserved but also 
\textit{constrained} to be an integer multiple of the electron charge $e$. Thus, 
if we make the simplifying assumption that there are no fractional excitations trapped in the bulk of the FTIs, we come to the conclusion that all $q_\mathsf{a}$'s must be integer-valued. This requirement restricts the possible eigenvalues of $\e^{i\pi q_\mathsf{a}}$ to $\pm 1$, corresponding to the even or odd fermion parity sectors. Without loss of generality, we will assume that each FTI has an even number of electrons, $q_{\sf a} = 0, \pm 2, \pm4, \dots$, leading to the set of conditions
\begin{equation}\label{chargeconstraint}
\e^{i\pi q_{\sf a}}=\prod_{k=1}^M P_{\left\lbrace\mathsf{a},2k \right\rbrace,\left\lbrace\mathsf{a},2k+1 \right\rbrace}=1.
\end{equation}
This choice amounts to restricting the twisted boundary conditions 
\eqref{alpha_boundary} to the periodic case.

The constraint \eqref{chargeconstraint} is violated if a quasiparticle or a quasihole 
is introduced in the bulk of the FTI. Due to the incompressibility of the FTI liquid, this process is related to the presence of an additional flux quantum $\Phi_0=h/2e$ piercing the bulk FTI \cite{levin2009,levin2011}. Thus, we can translate the conservation of electric charge on the edge of the FTI into a conservation of the magnetic flux threaded through the bulk. The latter is measured by the Aharonov-Bohm phase of a quasiparticle performing a counter-clockwise loop along the edge of the FTI. Mathematically, the Aharonov-Bohm phase factor is given by the string product 
$\Sigma_\mathsf{a}$ of the tunneling operators along such loop,
\begin{equation}\label{stringoperators}
\Sigma_\mathsf{a}=\prod_{k=1}^M P_{\left\lbrace\mathsf{a},2k-1 \right\rbrace,\left\lbrace\mathsf{a},2k \right\rbrace}
\end{equation}
which obeys $[\Sigma_{\sf a}, \e^{i\pi q_{\sf a}}]=0$, $[\Sigma_{\sf a}, H]=0$ and $(\Sigma_{\sf a})^{2m}=1$ and has eigenvalues $\sigma_\mathsf{a}=\e^{i\pi n/m}$, $n=0,\dots, 2m-1$. 
One can derive from Eqs. \eqref{commrules} and \eqref{commrules2} the commutation rule
\begin{equation}
\Sigma_{\mathsf{a}}\alpha^\dagger_{\{\mathsf{a},j\}}=e^{-i\pi\,/m} \alpha^\dagger_{\{\mathsf{a},j\}} \Sigma_{\mathsf{a}} \;\;,
\end{equation}
which confirms that the operator $\alpha^\dagger_{\{\mathsf{a},j\}}$, creating a charge $e/m$ on the edge of the FTI, at the same time adds $-\pi/m$ to the Aharonov-Bohm phase.

Consistently with the constraint \eqref{chargeconstraint}, we consider as physical only the sector of the full Hilbert space in which
\begin{equation}\label{fluxconstraint}
\Sigma_{\sf a}=1\;,
\end{equation} 
so that the magnetic flux in each FTI must be a multiple of $4m \Phi_0$. 
Changing the fluxes that pierce each FTI,  we choose a different set of eigenvalues 
$\sigma_{\sf a}$ and select a different physical sector.

The differences between the tile and stripe model originate from the fact that, for 
fixed system size, the number of FTIs is greater in the tile than in stripe model. This in turn determines the number of independent constraints on the Hilbert space, as reflected in the extent \(M\) of the product defining the string operator in 
Eq. \eqref{stringoperators} - we recall that $M=4$ in the tile model and $M=L_x$ in the stripe model. The result
is a different dimensionality of the physical sector of the Hilbert space of the two models.

\section{Mapping to 2D quantum clock models} \label{secJW}

To highlight the differences between the two models, it helps to use a mapping onto quantum 
2D $2m$-clock models. These models are defined on a 2D lattice where each site $\r$ is a 
$2m$-level quantum system and their Hamiltonians possess discrete $\mathbb{Z}_{2m}$ 
local symmetries. Clock Hamiltonians are defined in terms of degrees of freedom
$\sigma_\r$ and $\tau_\r$ that satisfy
\begin{align}
&\sigma_\r^{2m}=\tau_\r^{2m}=1\\
&\tau_\r^\dagger=\tau_\r^{-1}\;\;,\;\;\sigma_\r^\dagger=\sigma_\r^{-1}\ .
\end{align} 
Operators on any given site have commutation rules similar to those of PFs,
\begin{equation}
\sigma_\r\tau_\r=\e^{i\pi/m}\tau_\r\sigma_\r
\end{equation}
but operators on different sites commute. If \(m=1\) these relations are satisfied by Pauli matrices \(\sigma^z_\r, \sigma^x_\r\).
In the mathematical literature, the  algebra describing PFs is known as a generalized 
Clifford algebra. Its representation theory has been worked out in detail in 
Ref. \onlinecite{smith1991}, and from it, it is possible to infer a mapping  relating 
the PF operators $\alpha$ and the clock operators $\sigma, \tau$. This mapping can be 
achieved via a parafermionic Jordan-Wigner transformation\cite{fendley2012,smith1991,jagannathan2010}.
 
\begin{figure}[t!]
\includegraphics[width=0.8\columnwidth]{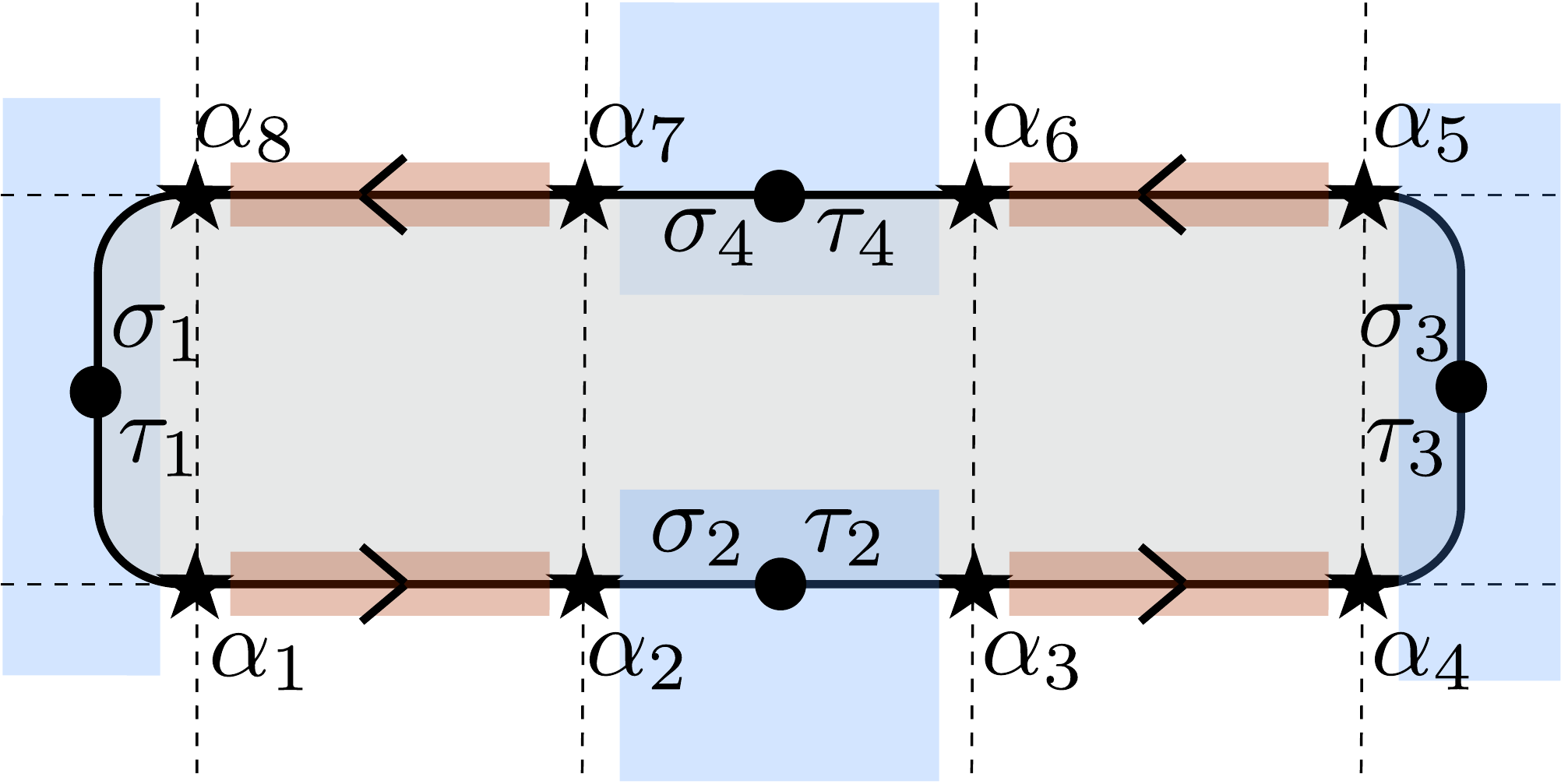}
\caption{The transformation defined in Eqs. \eqref{JW1}, \eqref{JW2} maps the PF operators living on the sites of a square lattice into a set of clock operators $\sigma,\tau$ defined on the links of the lattice occupied by a superconductor (marked in figure as black dots). Here, as an example, we show the positions of the clock operators in the case of a FTI in the tile model. The mapping between $\alpha_1\,\dots,\alpha_8$ and $\sigma_1, \tau_1, \dots, \sigma_4,\tau_4$ shown in this figure is explicitly written down in Eq. \eqref{tile8}.}
\label{fig_JW}
\end{figure}
 
\begin{figure*}[t!]
\centering
\includegraphics[width=\textwidth]{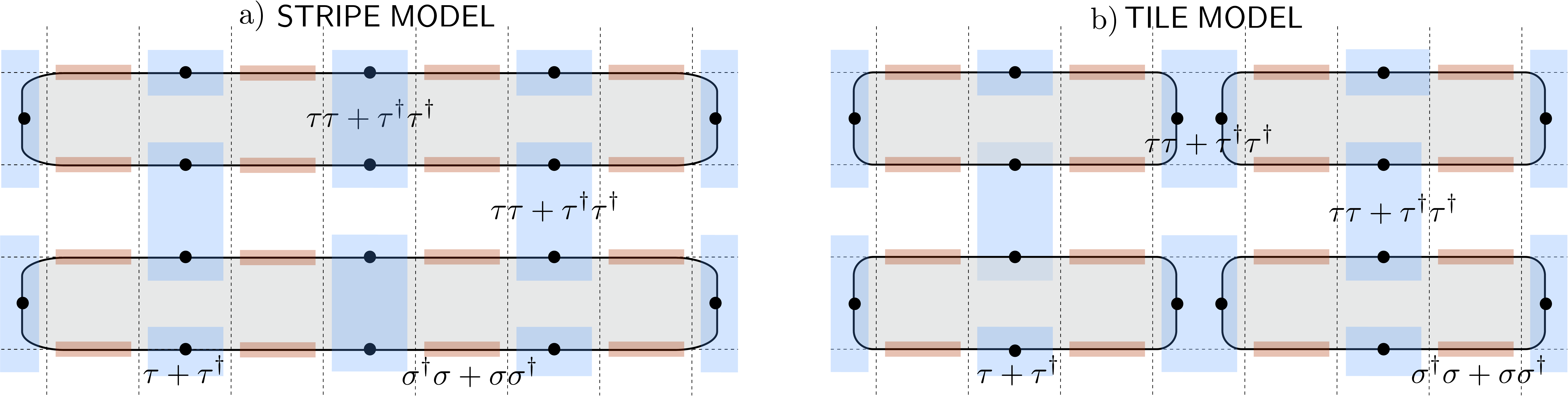}
\caption{Layout of the stripe and tile architecture in terms of clock operators $\sigma, \tau$ (black dots), living on the links of the square lattice occupied by a superconductor. Since clock operators at different sites commute, it is not necessary to order the FTI nor to assign an orientation to the FTI edges. However, notice that the clock operators for the two models live on two inequivalent lattices.}
\label{fig_clock_layout}
\end{figure*}
 
To each couple of adjacent PFs $\alpha_{\left\lbrace {\sf a}, 2k\right\rbrace },\alpha_{\left\lbrace {\sf a},2k+1\right\rbrace }$ on the same superconductor, 
we associate a couple of operators $\sigma_{{\sf a}, k}, \tau_{{\sf a}, k}$. These 
operators therefore live on (a subset of) the links of the square lattice defined by the 
PFs. The mapping between PFs and clock operators is given by
\begin{align}
\alpha_{\left\lbrace {\sf a},{2k}\right\rbrace }&=\kappa_{\sf a}\,
\sigma_{{\sf a},k+1}\,\prod_{1\le l\le k}\tau_{{\sf a},l} \label{JW1}\\ 
\alpha_{\left\lbrace {\sf a},{2k+1}\right\rbrace }&=\kappa_{\sf a}\,
\e^{i\frac{\pi}{2m}}\,\tau_{{\sf a},k+1}\,\sigma_{{\sf a},k+1}\,
\prod_{1\le l\le k}\tau_{{\sf a},l} \label{JW2}
\end{align} 
Here $\kappa_{\sf a}$ are fractional Klein factors taking care of the commutation rules 
between parafermions on different edges\cite{guyon2002,kim2006},
\begin{align}
\kappa_\mathsf{a}^{-1}&=\kappa_\mathsf{a}^\dag \\
\kappa_{\sf a}\kappa_{\sf a'}&=\e^{i\sgn({\sf a}'-{\sf a})\pi/m}
\kappa_{\sf a'}\kappa_{\sf a} .
\end{align}
These commutation rules must be compared with Eqs. (\ref{commrules},\ref{commrules2},\ref{signs}). Apart from fixing 
the commutators, the Klein factors do not play a role and drop out from any quadratic operator considered in this paper. The boundary conditions \eqref{alpha_boundary} and the constraints \eqref{chargeconstraint},\eqref{fluxconstraint} are taken into account by setting
\begin{align}
\sigma_{{\sf a}, M+1}&=\sigma_{{\sf a}, 1} \label{sigmaconst}\\
\tau_{{\sf a}, M+1}&=\tau_{{\sf a}, 1}     \label{tauconst}\\
\prod_{k=1}^M\,\tau_{{\sf a},k}&=1         \label{tauconst2}
\end{align}

Let us write down an explicit example of the transformation for the case $M=4$, relevant for the tile architecture. In this case the relations \eqref{JW1} and \eqref{JW2}, dropping the index $\mathsf{a}$ and the Klein factors for clarity, read (see also Fig.~ \ref{fig_JW})
\begin{equation}\label{tile8}
\begin{array}{ll}
\alpha_8 = \sigma_1 \,, & \alpha_1 = 
e^{i\frac{\pi}{2m}}\tau_1 \sigma_1\,, \\
\alpha_2 = \sigma_2 \tau_1\,, & \alpha_3 = e^{i\frac{\pi}{2m}}\tau_2 \sigma_2 \tau_1\,, \\
\alpha_4 = \sigma_3 \tau_2 \tau_1\,, & \alpha_5 = e^{i\frac{\pi}{2m}}\tau_3 \sigma_3 \tau_2 \tau_1\,, \\
\alpha_6 = \sigma_4 \tau_3 \tau_2 \tau_1\,,\;\;\; & \alpha_7 = e^{i\frac{\pi}{2m}}\tau_4 \sigma_4 \tau_3 \tau_2 \tau_1\, . \\
\end{array}
\end{equation}

We now rewrite the Hamiltonian in terms of the clock operators. For the Josephson energy, Eq. \eqref{JJham}, we obtain
\begin{equation}
 H_{\sf J}= -\frac{J}{2}\left(\e^{i\delta/2m}\,\sigma^\dag_{\mathsf{a},k+1}
\sigma_{\mathsf{a},k} + {\rm h.c.}\right)\;,
\end{equation}
while the charging energy, Eq. \eqref{HD}, becomes
\begin{equation}
H_\Delta=\begin{cases}-\dfrac{\Delta}{2}\,\left(\e^{i\pi n_{\sf ind}}\tau_{{\sf a},k}\,
+\rm{h.c.}\right)&\;\text{on the boundary} \\\\-\dfrac{\Delta}{2}\,
\left(\e^{i\pi n_{\sf ind}}\tau_{{\sf a},k}\tau_{{\sf a}',k'}+\rm{h.c.}\right)&\;
\text{in the bulk}
\end{cases}
\end{equation}
Note that the locality of the interactions is preserved. At this point, it is useful to split the array Hamiltonian of Eq. \eqref{Heff}
into bulk and boundary contributions,
\begin{equation}
H=H_{\sf bulk}+H_{\sf boundary},
\end{equation}
with
\begin{equation}\label{Hbulk}
H_{\sf bulk}=-\Big[\frac{J}{2}\sum_{\sf junctions}\sigma^\dagger_{{\sf a}, k+1}
\sigma_{{\sf a}, k}+\frac{\Delta}{2}\sum_{\substack{{\sf islands}\\\in\,\sf{bulk}}}
\tau_{{\sf a},k}\tau_{{\sf a}',k'}\Big]+\rm{h.c.}
\end{equation}
and
\begin{equation}\label{Hbdr}
H_{\sf boundary}=-\frac{\Delta}{2}\sum_{\substack{{\sf islands}\\\in\,\sf{bdr}}} 
\,\left( \tau_{\sf a, k}+\tau_{\sf a, k}^\dagger\right)\;,
\end{equation}
see also Fig.~ \ref{fig_clock_layout}. 
In writing Eqs. \eqref{Hbulk},\eqref{Hbdr} we have, for simplicity, set
$n_{\sf ind}=0$ for all islands and, in agreement with our choice of the physical sector, $\delta=0$ for all junctions.
Then the couplings \(J,\Delta\) become purely real and all equal.

Splitting the Hamiltonian into bulk and boundary contributions is useful for
studying various boundary conditions. For simplicity, 
in the remainder of this paper we will set $H_{\sf boundary} =0$ and focus on 
bulk properties of the array, assuming the system size is large enough to justify 
neglecting \(H_{\sf boundary}\). Also $H_{\sf boundary} =0$ corresponds
(for any system size) to the exact boundary conditions in case 
every superconducting island at the boundary of the array 
is grounded (since in that case $\Delta=0$ at the boundary). 

The Hamiltonian
\(H_{\sf bulk}\) of Eq. \eqref{Hbulk} commutes with an extensive set of local 
operators
\begin{equation} \label{xi}
\xi_{s}=\sigma_{\mathsf{a},k}\sigma_{\mathsf{a}',k'}^\dag \, , 
\qquad \left[ \xi_s , H \right]=0,  
\end{equation}
associated to every bulk superconducting island $s$ for both the tile and  
stripe models. Notice however that only those operators \(\xi_{s}\) that commute
with the constraints described in the previous section are actual 
physical symmetries: if $\mathsf{a}\neq\mathsf{a}'$, 
the operator \eqref{xi} moves one fractional charge from one edge to the other, 
thus violating the charge constraint. 

The difference between the tile and the stripe architectures can now be better appreciated, as shown in 
Fig.~ \ref{fig_clock_layout}. The 
effective Hamiltonian \eqref{Heff} associated to either 
architecture is defined on the square lattice in terms of PFs. In contrast,
\(H_{\sf eff}\) is defined on inequivalent lattices
when represented in terms of clock operators as in Eq. \eqref{Hbulk}. 
The tile model Hamiltonian \(H_{\sf tile}\) is obtained by specializing  
\(H_{\sf bulk}\) to a a decorated square lattice, while 
the stripe model Hamiltonian \(H_{\sf stripe}\) is obtained by 
specializing \(H_{\sf bulk}\) to a brickwall lattice.

\section{Topological phases and orders}
\label{secphase}

\begin{figure}[t!]
\includegraphics[width=\linewidth]{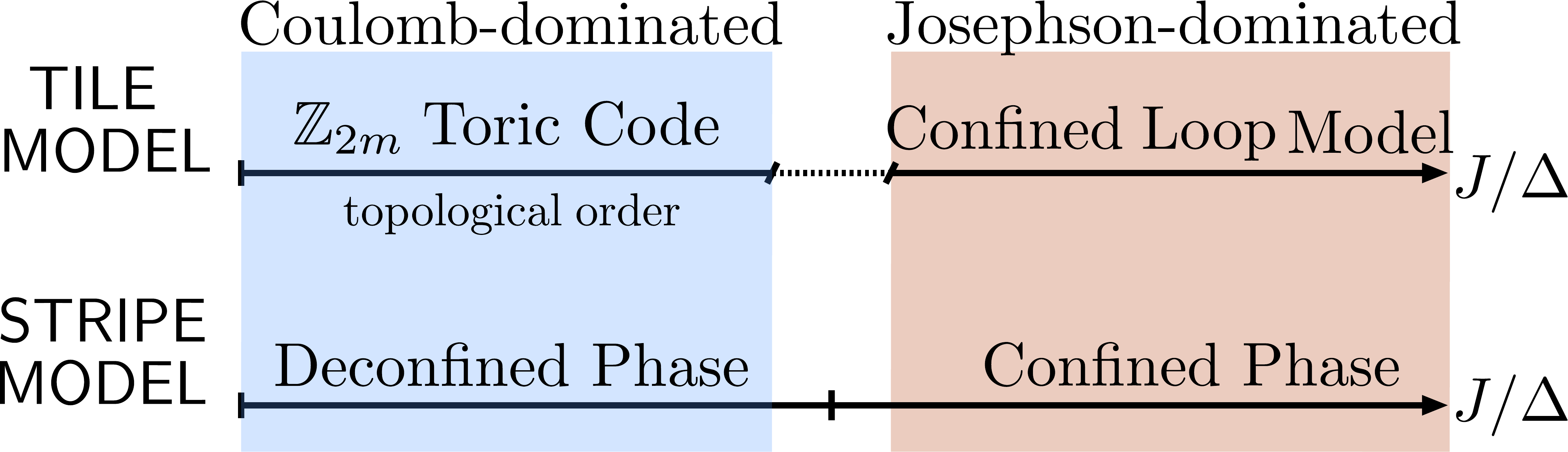}
\caption{Sketch of the phase diagram of the two models, as outlined in the introduction to Sec.~\ref{secphase}. We distinguish between two regimes, depending on whether Coulomb or Josephson energy dominates. The Coulomb regime shows, for both models, a non-local behavior, with degenerate ground states distinguished by the expectation values of string-operators. Only the tile model, however, presents a truly topological order characterized by anyonic excitations (see Sec.~\ref{sectoric}). The stripe model is instead dual to a $\mathbb{Z}_{2m}$ lattice gauge theory (see Sec.~\ref{secstripe}) which undergoes a deconfinement/confinement phase transition with increasing $J/\Delta$\cite{horn}. }
\label{fig_new_phasediagram}
\end{figure}

The quantum phase diagram of the tile and stripe models at zero temperature is controlled by 
the single parameter $J/\Delta$. In the following we will call {\it Coulomb-dominated} the regime $\Delta\gg J$ and {\it Josephson-dominated} the opposite regime $J\gg\Delta$. In this section we study the two regimes for both models, with a focus on the presence (or absence) of topological order. We dedicate Sec.~\ref{sectoric} to the tile model and Sec.~\ref{secstripe} to the stripe model. Let us summarize, here and in Fig.~ \ref{fig_new_phasediagram}, the main findings.

The Josephson-dominated regime shows no
topological features for either model. On one hand, the 
ground state of the tile model is singly degenerate in this limit due to the charge constraints.
Moreover, exactly at \(\Delta=0\), the ground wave
function reduces to a product state of wave functions for
local four-body clusters, emphasizing the absence of long-range entanglement. 
On the other hand, at \(\Delta=0\) the stripe model 
reduces to a system of decoupled, one-dimensional vector Potts chains
in zero transverse field. Hence, in the thermodynamic limit,
the stripe model has ferromagnetic order
in the Josephson-dominated regime $J\gg\Delta$.
The charge constraints do not suffice to
select a unique ground state like for the tile model, 
but rather correlates the magnetization for pairs of chains.  
 
In the opposite Coulomb-dominated regime, and specifically at \(J=0\),
both the tile and stripe models show dramatically increased (relative to 
\(\Delta=0\)) ground-manifold degeneracy. This suggests that at least one phase 
transition separates the two regimes in the thermodynamic limit for both models. 

We will show that the tile model is topologically ordered in 
the Coulomb-dominated regime, since 
\begin{enumerate}
\item the degeneracy of its ground manifold depends on the topology of the 
lattice, and
\item
the model has anyonic excitations completely equivalent to those in the
qudit toric code with \(\Z_{2m}\) discrete symmetry\cite{schulz2012,bullock2007}. 
\end{enumerate}
The second point is especially noteworthy, since the tile model is akin but
neither strictly equivalent to the \(\Z_{2m}\) toric code by Kitaev \cite{kitaev2003} nor to its generalizations \cite{bullock2007,schulz2012}. 

Unlike the tile model, which can be defined naturally on a surface of 
arbitrary genus due to the limited extension of its FTIs,
 the stripe model fits naturally only open, cylindrical, 
or periodic (toroidal) boundary conditions. We will see that in the Coulomb-dominated regime its ground state degeneracy is not protected against local operators. Hence we do not consider
the stripe model topologically ordered. We will argue nevertheless that the $\Delta\gg J$ regime is characterized by a non-local order parameter, which we will define using a duality mapping the stripe model to the \(\Z_{2m}\) lattice gauge theory \cite{horn}. Thus, even in the absence of a topological order, the Coulomb-dominated phase of the stripe model can be addressed more generically as a topological phase.

The connection between topological order and lattice gauge theories in Josephson junction arrays has already been the subject of detailed studies, for both $\mathbb{Z}_2$ symmetric models \cite{ioffe2002,ioffe2002b,doucot2002} and more general Abelian and non-Abelian gauge symmetries \cite{doucot2004}. These models are based on superconducting architectures of Josephson junctions, where the required degeneracies are obtained with fine tuned magnetic fluxes. Such architectures can present topological phases in the Josephson-dominated regime, as experimentally verified in Ref. \onlinecite{gladchenko2009}. The two models studied differ in two important aspects: the absence of fine-tuning to create and control the elementary components of the arrays, which is due to the topological origin of the PF modes, and the fact that the topological phases are obtained in the Coulomb-dominated regime, essentially exchanging the role of electric and magnetic excitations with respect to Ref. \onlinecite{ioffe2002}.

\subsection{A physical realization of $\bf \mathbb{Z}_{2m}$ toric code anyons: the
tile model} \label{sectoric}

\begin{figure}[t!]
\includegraphics[width=\columnwidth]{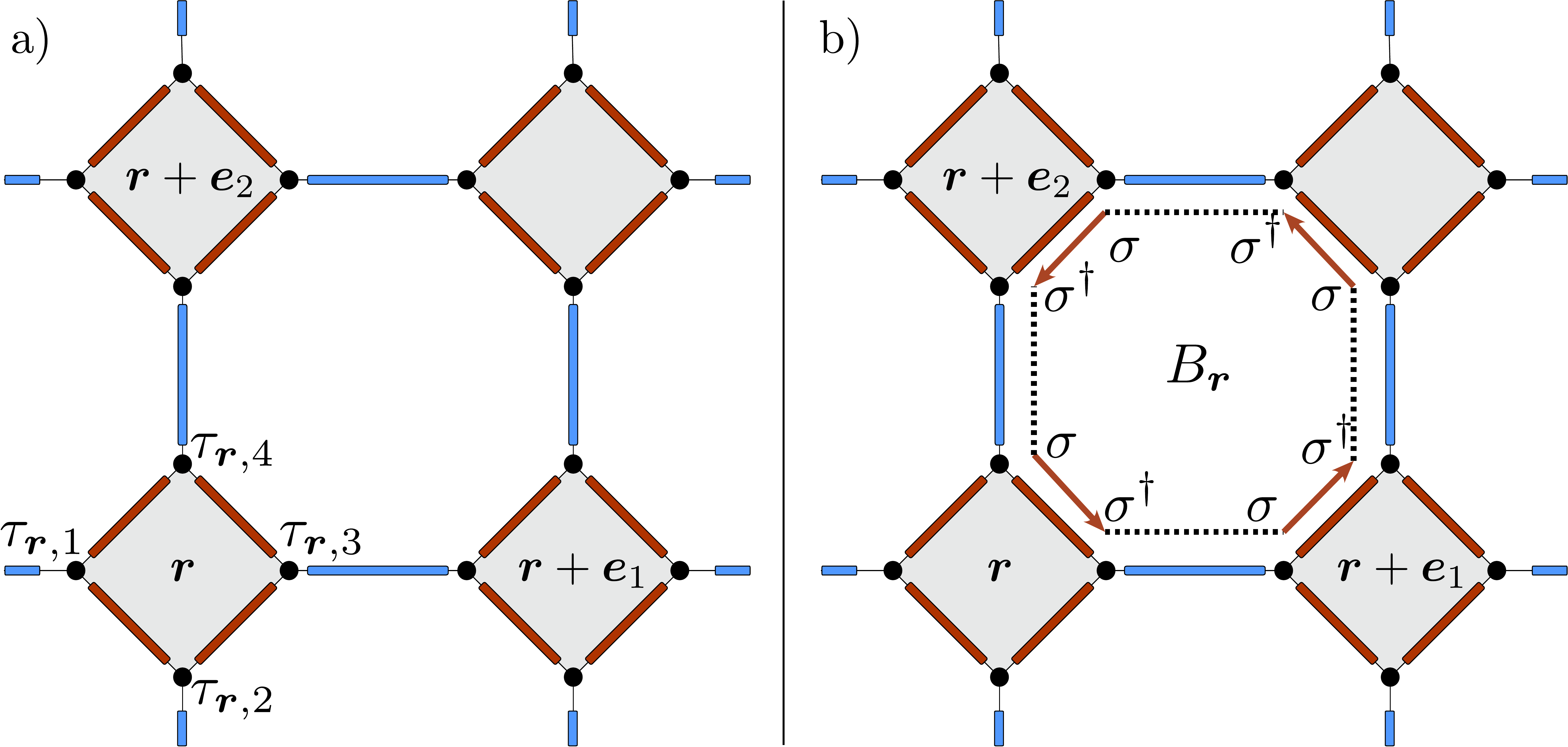}
\caption{Panel (a): The notation adopted for the decorated square lattice on which the tile model can be conveniently rearranged. The grey diamonds, sitting on the sites $\r$ of a square lattice, are FTI. To each site $\r$ there correspond four clock operators $\sigma_{\r,i}, \tau_{\r, i}$, arranged counterclockwise. Blue links are SC, red links are FM. Panel (b): The operator $B_\r$ in Eq. \eqref{plaquetteoperator} is the counterclowise product of four $\sigma\sigma^\dagger$ operators around the same plaquette.\label{fig_tile_lattice}}
\end{figure}

To analyze the effective Hamiltonian for the tile architecture, 
it is useful to adopt a decorated square lattice were 
each FTI sits on a site $\r=i\i+j\j$, with $(i,j)$ a pair of integers, see 
Fig.~ \ref{fig_tile_lattice}. In this lattice, the array Hamiltonian \(H_{\sf bulk}\)
of Eq. \eqref{Hbulk} is given by
\begin{align}
H_{\sf tile}=&-\Big[\frac{J}{2}\sum_\r\sum_{i=1}^4\sigma^\dagger_{\r,i}
\sigma_{\r,i+1}+\frac
{\Delta}{2}\sum_{\left\langle\r,\r' \right\rangle} Q_{\left\langle \r,\r'\right\rangle }
\Big]+{\rm h.c.}
\end{align}
where $Q_{\left\langle \r,\r'\right\rangle }$ labels the charging energy terms of
the superconducting islands, now sitting on the links of the square lattice between 
two neighboring diamonds, 
\begin{align}
Q_{\left\langle \r,\r+\i\right\rangle }=\tau_{\r,3}\tau_{\r+\i,1}\,,
\quad Q_{\left\langle \r,\r+\j\right\rangle }= \tau_{\r,4}\tau_{\r+\j,2}. \label{link}
\end{align}

Let us now consider the limit $J=0$ deep in the Coulomb-dominated regime. 
The system is then in a limit state where tunneling between islands is forbidden. Each superconductor minimizes the charging energy in the space of physical states 
specified by the charge constraints:
\begin{equation}
Q_{\left\langle \r, \r+\i\right\rangle }=Q_{\left\langle \r, \r+\j\right\rangle }=1.
\end{equation} 
These conditions allow for $(2m)$-fold degeneracy for each superconducting link, corresponding to the presence of the local symmetries \eqref{xi}. However in the sector of physical states we must impose the constraints 
\begin{equation}
\prod_{i=1}^4 \tau_{\r, i}=1
\end{equation}
derived from Eq. \eqref{chargeconstraint}. Nearly half of the previous states are then 
projected out, leaving a ground state manifold of dimension 
$(2m)^{\# {\sf SC}/2}$. This number is exact asymptotically
in the system size, but depends slightly on the boundary conditions. 
For example, for periodic boundary conditions the {\it exact} degeneracy 
of the ground manifold is $(2m)^{1+(\# {\sf SC}/2)}$.

The degeneracy of the Coulomb-dominated limit at $J=0$ is partially 
lifted when weak tunneling terms are reintroduced, that is, we allow
$J\neq 0$. The Coulomb-dominated regime $J\ll\Delta$ can be 
treated perturbatively by introducing an effective low-energy Hamiltonian affecting only the ground state manifold at $J=0$. We need to keep only those operators in the perturbative expansion that do not couple the ground state manifold to the excited states. This is a standard technique \cite{bravyi2011}, and the computation is 
analogous to the perturbative derivation of the $\Z_2$ toric code Hamiltonian from 
Kitaev's honeycomb model \cite{kitaev2006}, so we will only streamline the essential points.

At first order, the perturbation $\sigma^\dagger_{\r,i}\sigma_{\r,i+1}$ creates 
two charged $\pm e/m$ excitations on adjacent superconductors, increasing the energy 
of the system by an amount $G=2\Delta(1-\cos\pi/m)$. 
Similarly, at all odd orders we obtain terms that we neglect as they do not 
leave the \(J=0\) ground state manifold invariant. At second order, we obtain 
only terms describing the tunneling back and forth of a fractional charge $e/m$ 
across a single Josephson link. 
These terms renormalize the ground energy level, that is, they
provide an energy offset to the full Hamiltonian. At fourth order we obtain the first 
relevant contribution. 
It is a plaquette operator of the form (see Fig.~ \ref{fig_tile_lattice})
\begin{align}\label{plaquetteoperator}
B_\r=&\left(\sigma_{\r,4}\,\sigma^\dagger_{\r, 3}\right)\times\left(\sigma_{\r+\i,1}\sigma_{\r+\i,4}^\dagger\right)\times\\
&\times\left(\sigma_{\r+\i+\j,2}\,\sigma^\dagger_{\r+\i+\j,1}\right)\times\left(\sigma_{\r+\j,3}\sigma_{\r+\j,2}^\dagger\right)\notag,
\end{align}
describing the tunneling of an $e/m$ excitation along a loop of four FTI edges and 
four superconducting islands. The resulting perturbative Hamiltonian reads
\begin{equation}
H_{\sf pert}^{\sf tile} = -\Big[\frac{\Delta}{2}\sum_{\left\langle\r,\r' \right\rangle} 
Q_{\left\langle\r,\r' \right\rangle} + 
\frac{5J^4}{4G^3} \sum_{\r}  B_\r \Big]+  {\rm h.c.}
\end{equation}
where we note that, in the case $m=1$, the coefficient of $B_\r$ matches the one obtained in a similar perturbative expansion in Ref. \onlinecite{terhal2012}, where equivalent plaquette operators are obtained.

Since the operators \(B_\r\) commute with the charge constraints,
the space of physical states for the perturbative Hamiltonian \(H_{\sf pert}^{\sf tile}\)
is left untouched. The bond operators $Q_{\left\langle\r,\r' \right\rangle}$ and $B_\r$, together with their Hermitian conjugates, constitute a completely commuting set of stabilizers for a qudit surface code\cite{bullock2007}.
This surface code protects against every local error that excites a ground state of \(H_{\sf pert}\)
into a state of higher energy\cite{bullock2007}.
In particular, the \(Q_{\left\langle\r,\r' \right\rangle}\) operators enforce the absence of 
charge excitations in the superconducting islands, while the \(B_\r\) operators 
enforce the absence of flux excitations.

Let us note that the Hamiltonian \(H_{\sf pert}^{\sf tile}\) is not exactly equivalent to the $\mathbb{Z}_{2m}$ toric code originally discussed by Kitaev in Ref. \onlinecite{kitaev2003}, since the stabilizers $Q_{\left\langle\r,\r' \right\rangle}$ and $B_\r$ are not projectors. The construction of these operators is instead more closely related to the qudit surface codes introduced in Ref. \onlinecite{bullock2007} - although, strictly speaking, \(H_{\sf pert}^{\sf tile}\) is not equivalent to those models as well, since it is not possible to canonically associate our stabilizers
$Q_{\left\langle\r,\r' \right\rangle}$ and $B_\r$ to vertices and faces of a two-dimensional simplicial complex. Despite these minor differences, however, the topological properties of these models are the same.

\begin{figure}[t!]
\includegraphics[width=\columnwidth]{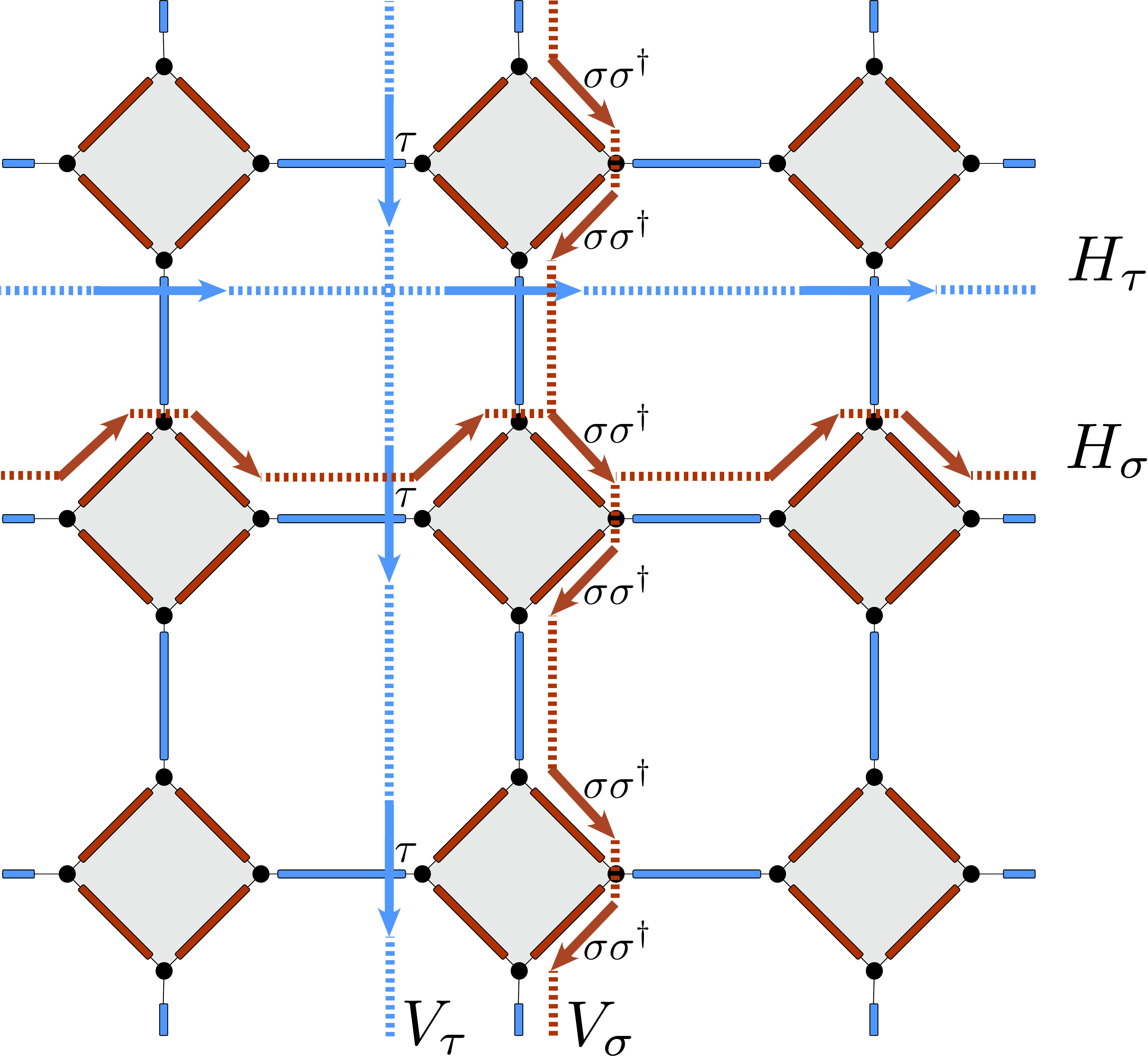}
\caption{The four non trivial loop operators that define the ground state manifold of the tile model on a torus. The operators $H_\tau, V_\tau$ are defined as the product of all $\tau$ along the path described by the two blue lines, in the order established by the arrows. Similarly, the operators $H_\sigma, V_\sigma$ are defined as the product of all $\sigma\sigma^\dagger$ operators along the path given by the red lines. Loop operators corresponding to different paths only differ by a product of stabilizer operators $Q_{\left\langle \r,\r'\right\rangle}$ or $B_\r$.\label{fig_torus_loops}}
\end{figure}

The conditions 
\begin{equation} \label{stabilizers}
Q_{\left\langle\r,\r' \right\rangle}=1,\qquad B_\r=1, \qquad \prod_i\tau_{\r,i}=1,
\end{equation}
specify the ground manifold of the system. 
Its degeneracy can be determined from symmetry considerations. 
On the torus (periodic boundary conditions) there are four types of
loop symmetries $H_\tau, H_\sigma, V_\tau, V_\sigma$, defined in 
Fig.~ \ref{fig_torus_loops}, associated to non-contractible loops
and compatible (commuting) with the charge constraints.
It is interesting to notice (for comparison with other \(\Z_{2m}\) surface codes
in the literature) that the loop symmetries 
\(H_\tau,V_\tau\), (\(H_\sigma, V_\sigma\)) are disjoint, that is, they do not 
have any clock degrees of freedom in common.
As usual, any two loop symmetries 
of a given type, \(\tau\) or \(\sigma\),
associated to equivalent but different non-contractible loops
differ only by a product of the stabilizer operators in (\ref{stabilizers}) (or their hermitian conjugate). Hence, in the
ground manifold, all these loop symmetries collapse into just four
inequivalent ones. These form two non-commuting pairs,
\begin{align}
H_\tau\,V_\sigma&=\e^{-i\pi/m}\,V_\sigma\,H_\tau,\\
H_\sigma\,V_\tau&=\e^{-i\pi/m}\,V_\tau\,H_\sigma,
\end{align}
while $[H_\tau, H_\sigma]=[V_\tau, V_\sigma]=[V_\tau, V_\sigma]=[H_\tau, H_\sigma]=0$. 
Since $V_\sigma^{2m}=\mathds{1}=H_{\sigma}^{2m}$, it follows that each pair 
identifies $2m$ different ground states, yielding a ground state degeneracy of 
$(2m)^2$. This is the dimension of the code space defined by \(H_{\sf pert}^{\sf tile}\).

Our stabilizer code can be adapted to a planar geometry with open boundary conditions
along the lines set in Ref. \onlinecite{bravyi1998}. In this case there will be
only two non-trivial string operators and thus $2m$ ground states. 
In planar geometries with $g$ holes, the ground-state manifold degeneracy 
increases to $(2m)^g$.

\begin{figure}[t!]
\includegraphics[width=\columnwidth]{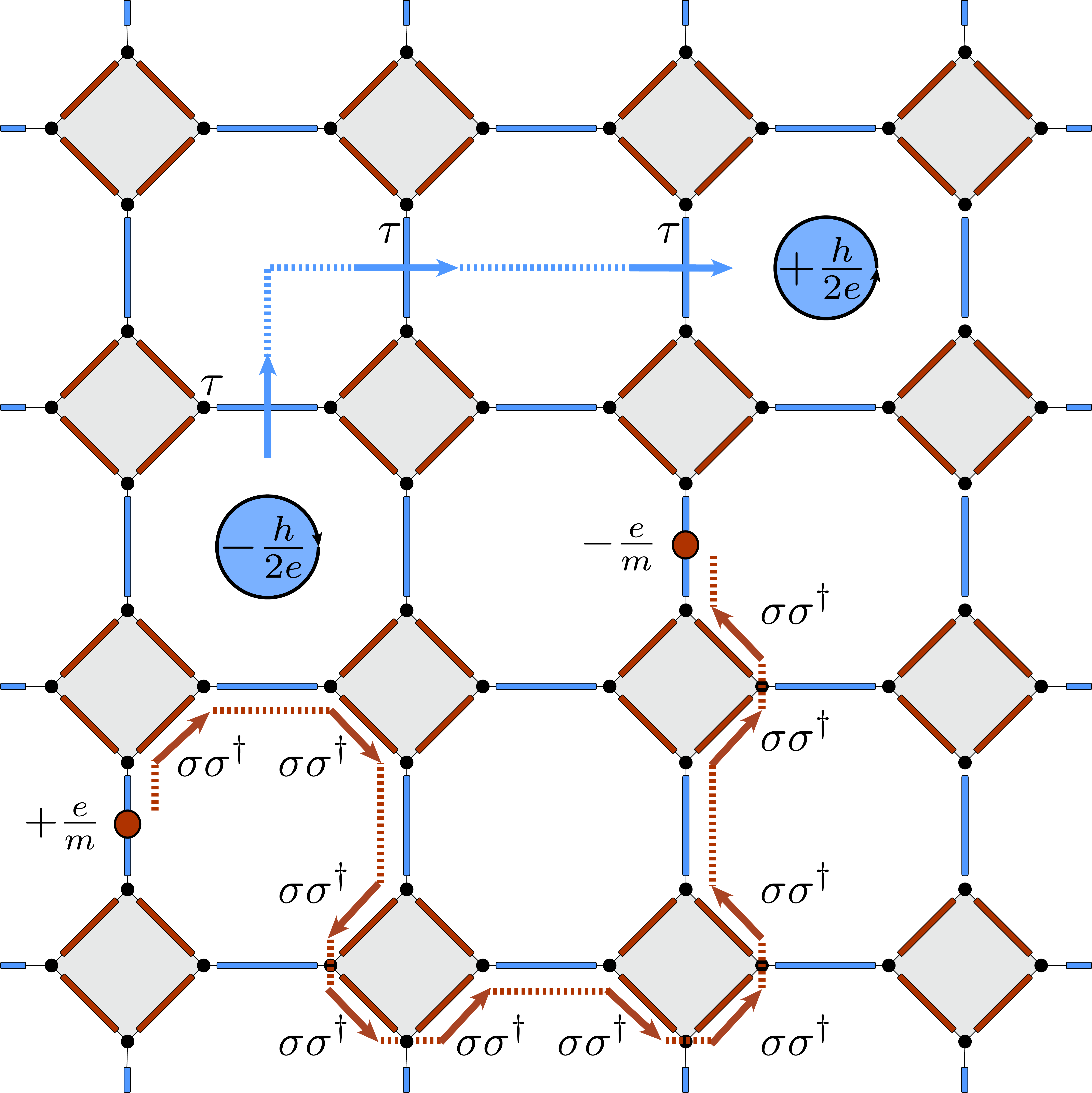}
\caption{Vortex and charge excitations in the tile model, appearing at the ends of open string of $\tau$ and $\sigma\sigma^\dagger$ operators respectively. The vortices live on the plaquettes of the lattice, while the charges on the superconducting links. They are mutual Abelian anyons, with an exchange phase $\e^{i\pi/2m}$.\label{fig_tile_excitations}}
\end{figure}

The Hamiltonian \(H_{\sf pert}^{\sf tile}\) has two different types of excitations
illustrated in Fig.~ \ref{fig_tile_excitations}:
\begin{enumerate}
	\item Two localized charge excitations $\pm e/m$ can be created on two different links by an open string of tunneling operator of the form $S=\prod(\sigma\sigma^\dagger)$. The operator $S^\dagger$ switches the sign of the charges at the end of the open string.
	\item Two $\pm h/2e$ vortices are created on neighboring plaquettes by one of the two operators $\tau$ on the link separating the plaquettes. (The other $\tau$ operator belonging to the same link creates the same vortices, but with opposite sign.) The vortices  can be moved apart without further energy costs applying a string $T=\prod \tau$ of consecutive $\tau$ operators sharing one common plaquette.
\end{enumerate}
Both charge and flux excitations are bosons when considered separately (since different $S$ operators commute with themselves, as well as different $T$ operators). However, when a charge excitation is moved in a loop around a flux excitation, the wavefunction will acquire a (Aharonov-Bohm) phase $\e^{i\pi/m}$, implying that charges and flux excitation are mutually Abelian anyons with a fractional exchange phase $\e^{i\pi/2m}$. This can be verified by computing the commutator of a pair of $S$ and $T$ strings intersecting each other. Additionally, the underlying $\Z_{2m}$ symmetry allows the presence of multiple excitations of charge $ne/m$ and flux $nh/2e$, with $n=0,\dots,2m-1$, created by the $n$-th power of $S$ and $T$ operators, as in the usual qudit surface codes \cite{bullock2007}.

Let us discuss possible terms that may destroy 
the topological order. Higher orders in perturbation theory yield larger loop operators, which can be decomposed in terms of products of $B_\r$ operators and their powers. These higher-order terms commute with $H_{\sf pert}^{\sf tile}$ and strengthen the absence of fluxes in the plaquettes, leaving the ground-state manifold intact. The description breaks down only when the perturbation order $L$ is equal to the system size. At this point, the loop operators $H_\sigma, V_\sigma$ are generated in the perturbative expansion, lifting the ground-state degeneracy by an energy $O(J^L/\Delta^{L-1})$.

However, we may worry about external perturbations of the form 
\begin{equation}
h\sum_{\r, i}\,(\tau_{\r,i}+\tau^\dagger_{\r, i}),
\end{equation}
which would break the ground state degeneracy. This perturbation corresponds to an external magnetic field in the vector Potts description and may drive a transition from the topologically ordered phase to a topologically trivial one constituted by a condensate of the vortex excitations. For the \(\Z_{3}\) toric code, this transition was 
observed numerically in Ref. \onlinecite{schulz2012}. 
A general analysis\cite{burnell2011,burnell2012} of the phase diagram of 
$\Z_p$ (\(p=2,3,\dots\)) Wen-Levin models\cite{levin2005} suggests that the 
transition, in the 2+1D transverse-field Potts universality class, is of the first 
order for any $p>3$ ($m>1$), thus easily detectable due to the discontinuity in 
the energy density. Our model however is of the vector Potts (rather than simple
Potts) type and further investigations are required to assert the equivalence of the two cases for generic $m$. 

Finally, let us briefly discuss the Josephson-dominated regime. The topological order in the Coulomb-dominated regime of the tile model disappears when the tunneling terms become comparable to the charging energy. 
In the opposite extreme limit, $\Delta=0$, the FTIs decouple
and the Hamiltonian is just the sum of the Josephson interactions along each diamond of 
the lattice in Fig \ref{fig_tile_lattice}. In particular, to minimize the energy, the four 
clock operators $\sigma$ for each FTI must be aligned and, considering the charge constraint (\ref{tauconst2}), one obtains that the ground state of each FTI is constituted by an equal superposition of all the polarizations:
\begin{equation}
 \ket{{\sf GS}}_\r = \frac{1}{\sqrt{2m}} \sum_{k=0}^{2m-1} \ket{\sigma_{\r,1}=\sigma_{\r,2}=\sigma_{\r,3}=\sigma_{\r,4}=e^{i\frac{k\pi}{m}}}.
\end{equation}
Thus the total ground state is simply the product of the states $\ket{{\sf GS}}_\r$ of 
all the FTIs. Due to the charge constraints, it is unique independently of the topology of the system. 

For fixed system size, the ground state will remain non-degenerate also when we consider a small
charging energy contribution, $\Delta\ll J$, in the Josephson-dominated regime. In particular the effect of applying all the charging operators $Q_{\left\langle\r,\r' \right\rangle}$ in a closed area $\mathcal{S}$ is to rotate all the clock operators $\sigma$ inside $\mathcal{S}$. The result is the formation of a domain wall constituted by all the links along the edge of $\mathcal{S}$, where the clock operators are not aligned anymore. The energy cost of the domain wall is proportional to $J\partial\mathcal{S}$, where $\partial\mathcal{S}$ is the number of broken Josephson links along the perimeter of $\mathcal{S}$. Therefore, the Hamiltonian in the Josephson dominated regime can be seen as the confined phase of a loop model\cite{fendley2008}, where the loops are the edges of domains with different spin alignment: $J$ provides a tension to the loops whereas $\Delta$ constitutes their kinetic energy. Between the topologically ordered Coulomb-dominated regime and the topologically trivial Josephson-dominated regime other phases may appear and the full phase diagram of the tile model deserves further investigations.  

\subsection{The stripe model and the \(\Z_{2m}\) Gauge theory}\label{secstripe}

As anticipated at the end of the previous section, the stripe model is naturally supported
on a brick-wall lattice.  It is convenient to place the clock degrees of 
freedom \(\sigma_{(i,j)}\) and \(\tau_{(i,j)}\) on the sites 
$$\{(i,j)\ |\ i=0,\dots,L_x-1,\quad j=0,\dots, L_y-1\}$$
of an \(L_x\times L_y\) square lattice and distinguish between the two sublattices defined by the conditions $(i+j)={\sf even}$ and $(i+j)={\sf odd}$. In the bulk of this geometry the stripe model becomes the generalization of the XXZ honeycomb compass model \cite{nussinov2012} with $\mathbb{Z}_{2m}$ symmetry:
\begin{multline}
H_{\sf stripe}= 
-\Big[\frac{\Delta}{2}\sum_{i+j={\sf even}} Q_{(i,j)}\\
+\frac{J}{2}\sum_{j=0}^{L_y-1}\sum_{i=0}^{L_x-2} 
 \sigma_{(i,j)}\sigma^\dagger_{(i+1,j)}\Big]+h.c.
\end{multline} 
with 
\begin{equation}
Q_{(i,j)}\equiv \tau_{(i,j)}\tau_{(i,j+1)}.
\end{equation}
Depending on the chosen boundary condition, $H_{\sf stripe}$ must be supplemented 
with an additional boundary term that we will disregard for the sake of simplicity.
The stripe unitary operators
\begin{equation}\label{stripe_const}
S_jS_{j+1}=\prod_{i=0}^{L_x-1} Q_{(i,j)}, \qquad j=0,2,\dots,L_y-2
\end{equation}
represent the physical constraint on the electric charge 
of the FTIs, Eq. \eqref{tauconst2}; therefore the physical states $\ket{\Psi}$ must satisfy:
\begin{equation}\label{stripes_const}
S_jS_{j+1}\ket{\Psi}=\ket{\Psi},\quad j=0,2,\cdots, L_y-2.
\end{equation}

\begin{figure}[t!]
\includegraphics[width=\columnwidth]{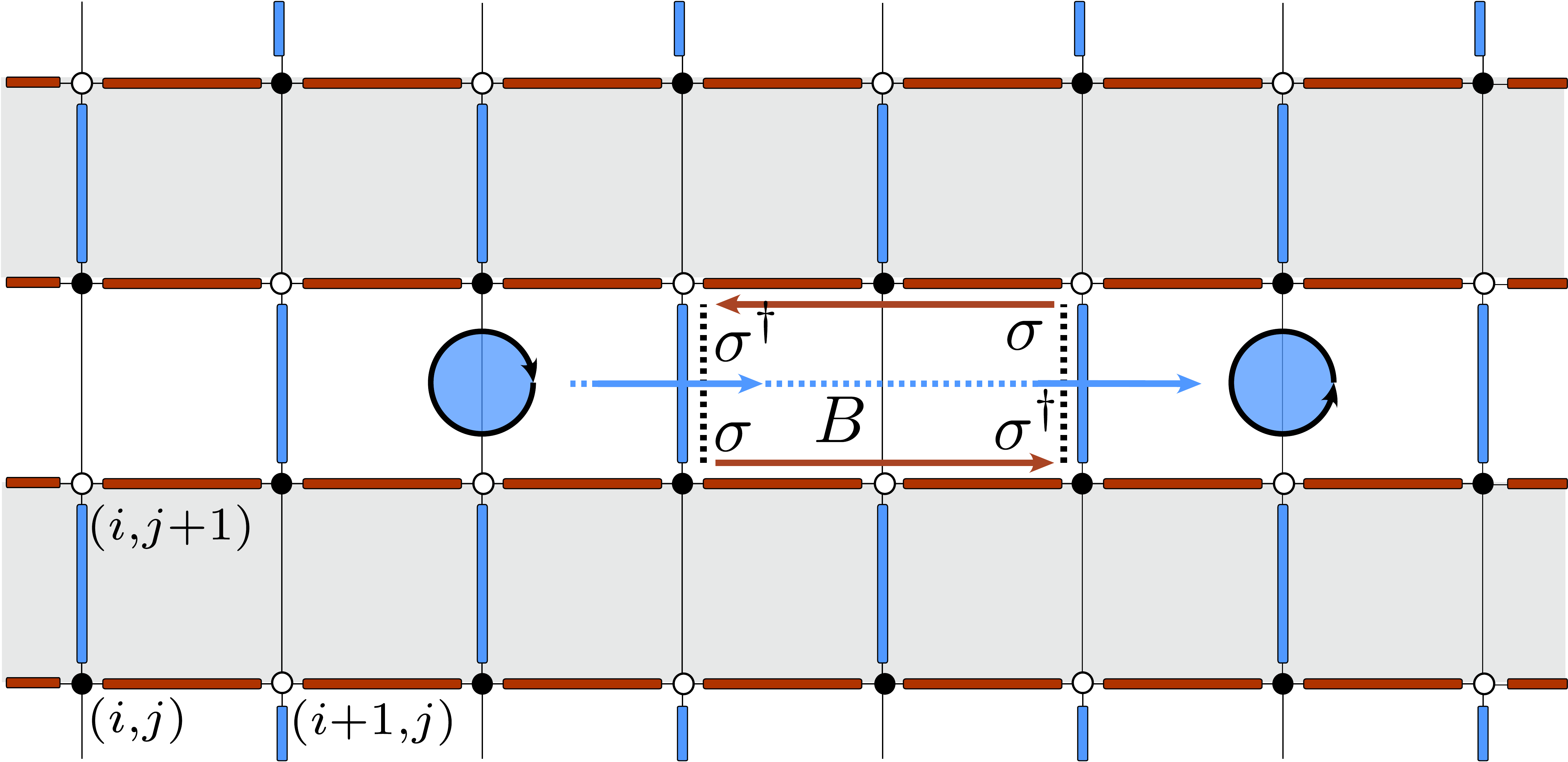}
\caption{Notation adopted to study the stripe model. The black and white sites identify clock operators $\sigma, \tau$, distinguished by a sublattice degree of freedom. Grey stripes are FTI, blue links are superconducors and red links ferromagnets. Notice that half of the vertical links are missing, thus the stripe model is effectively defined on a brick-wall lattice. The flux threaded through a single plaquette of the lattice is measured by the operator $B$ in the figure (see Eq. \eqref{upsilons}). As in the tile model, flux excitations (the oriented blue circles) can be created by an open string of $\tau$ operators: however, the geometry constrains their movement in the horizontal direction. \label{fig_stripe_lattice}}
\end{figure}

The next task is to specify the physical symmetries of the stripe model. 
The set of non-trivial unitary operators that commute with \(H_{\sf stripe}\) 
is generated by
\begin{eqnarray}
S_j&=&\prod_{i=0}^{L_x-1} \tau_{(i,j)},\qquad \qquad j=0,1,\dots, L_y-1,
\\
\xi_{(i,j)} &=& \sigma_{(i,j)}\sigma^\dagger_{(i,j+1)},\qquad i+j={\sf even}.
\end{eqnarray}
We need to specify those operators in this set that also commute with the charge
constraints of Eq. \eqref{stripes_const}. The symmetries 
\(S_j\) trivially satisfy this condition, but they are not all independent in the 
sector of physical states. We can keep 
\begin{equation}\label{1dsyms}
S_j,\qquad j=0,2,\cdots, L_y-2,
\end{equation}
as an independent set of one-dimensional symmetries for the stripe model.
As explained in the introduction to this section, these symmetries are spontaneously broken in the 
Josephson-dominated regime (at zero temperature). The effective dimensional reduction displayed by the
stripe model in this regime is intimately connected to the one-dimensional symmetries
of Eq. \eqref{1dsyms} \cite{dim_red}. 

The local physical symmetries of the stripe model are given by the minimal combination of the 
operators \(\xi_{(i,j)}\) that commute with the charge constraints and they assume the form
\begin{equation}\label{upsilons}
B_{(i,j)} = \xi_{(i,j)}\xi^\dag_{(i+2,j)},\qquad i+j={\sf even}.
\end{equation}
These local symmetries have an immediate interpretation: they
describe the Aharonov-Bohm phase associated to the 
magnetic fluxes threading the plaquettes of the brick-wall lattice (see Fig.~ \ref{fig_stripe_lattice}).

The stripe model has no global symmetries independent on the lower-dimensional 
symmetries already discussed. The global symmetry
\(
S_1S_3\dots S_{L_y-1}
\) is trivially spontaneously broken in the Josephson-dominated
regime by the spontaneous breakdown of its one-dimensional constituents. 
Other global symmetries appear as products of the local symmetries discussed in the
previous paragraph, and so cannot be spontaneously broken by Elitzur's theorem \cite{elitzur75}. 
This suggests that any ordered phase of the stripe model 
(outside the Josephson-dominated limit) must be characterized in terms of a 
generalized, non-local order parameter \cite{holographic}. However this is not enough to assert that the system shows topological order according to our previous definition based on the topological ground state degeneracy and the presence of anyonic excitation. Rather, in the absence of a Landau 
local order parameter, we speak more generically of topological phases. 

To the purpose of comparing the tile and stripe models in the 
Coulomb-dominated regime it is useful to perform a perturbative analysis also of the 
stripe model in the limit $\Delta \gg J$. Just as for the tile model, the first 
non-trivial term appears at the fourth order in perturbation theory, where the perturbative
Hamiltonian of the stripe model becomes:
\begin{equation}\label{pert_stripe}
H^{\sf pert}_{\sf stripe} = - \sum_{i+j={\sf even}}\Big[ 
\frac{\Delta}{2}Q_{(i,j)} + \frac{5J^4}{4G^3} B_{(i,j)}\Big]  
+ {\rm h. c.}
\end{equation}
As in the case of the tile model, also for this Hamiltonian it is 
possible to define localized fractional charge excitations and vortex excitations as end of open strings of $\tau$ and $\sigma\sigma^\dagger$ operators. However, for this architecture the vortex excitations can propagate only in the horizontal direction, as can be realized noting that superconductors of different rows share no common plaquette. 

Indeed, in the Hamiltonian $H^{\sf pert}_{\sf stripe}$ each row of superconducting islands is decoupled from  
the others, and is characterized by a ground state degeneracy of $2m$. However, the rows between different FTIs present a non-physical symmetry $\xi$ which does not 
commute with the constraint (\ref{stripes_const}). Thus, accounting for the 
charge constraints, the overall degeneracy of the 
ground states in the physical sector is $(2m)^{\#{\rm FTIs}}$. Crucially, this degeneracy is not protected against the local symmetries $\xi_{(i,j)}$. Since these local operators may cause transitions between different ground states, the stripe model does not posses a proper topological order. Despite this fact, the model is characterized by a 
non-local order parameter, as we will discuss in the following. 

To this purpose, and more in general to investigate the phase diagram, it is useful to exploit the bond-algebraic theory 
of dualities \cite{cobanera2011,conprl} which allows us
to study the bulk properties of the constrained stripe Hamiltonian for large system 
size. Our strategy will be to 
find a duality (consistent 
with the constraints), mapping \(H_{\sf stripe}\) to a known model. 
As shown in Refs. \onlinecite{conprl,cobanera2011}, quantum dualities
can be obtained as isomorphisms of bond algebras of interactions preserving locality.
In principle, we could study the minimal bond algebra of interactions 
generated by the bonds \(Q_{(i,j)}\ (i+j={\sf even})\) and
\(\sigma_{(i,j)}\sigma^\dagger_{(i+1,j)}\) in \(H_{\sf stripe}\). 
However, a duality derived from 
this bond algebra, that is, an alternative local representation of these 
interactions, may not preserve the charge constraints of Eq. \eqref{stripes_const}, 
because these constraints are not contained in this minimal bond algebra. 
Hence we consider a larger set of generators 
\begin{align}
& Q_{(i,j)}\;,&\!\!\! \quad i=0,\cdots, L_x-1\,;\quad j=0,\cdots, L_y-2;
\nonumber\\
& \sigma_{(i,j)}\sigma^\dagger_{(i+1,j)}\;,&\!\!\! \quad i=0,\cdots, L_x-2\,;\quad j=0,\cdots, L_y-1
\nonumber
\end{align}
(and Hermitian conjugates) for the stripe model's bond algebra 
\(\mathcal{A}_{\sf stripe}\). That is, we are including also the bonds  \(Q_{(i,j)}\ (i+j={\sf odd})\), which are
absent from the Hamiltonian. Such extended bond algebra does contain the charge constraints
in Eq. \eqref{stripes_const}; hence a duality for \(\mathcal{A}_{\sf stripe}\) maps these constraints in a
well defined fashion either to the identity operator (in which case the duality
solves the constraints \cite{cobanera2011}) or to constraints of the dual model. 

The characterization of \(\mathcal{A}_{\sf stripe}\) in terms of relations  
among its bond generators reveals the following dual representation of the bond algebra
of interactions:
\begin{eqnarray}
Q_{(i,j)}&\dual& B_{{\sf d} \,(i,j)},\label{d1}\\
\sigma_{(i,j)}\sigma^\dagger_{(i+1,j)}&\dual& \tau_{(i,j)},\label{d2}
\end{eqnarray}
with 
\begin{equation}\label{def_plaquette}
B_{{\sf d}\, (i,j)}\equiv\left\{
\begin{array}{ll}
\sigma^\dagger_{(i,j)}\sigma^\dagger_{(i,j+1)}& \mbox{if}\ i=0,\\
\sigma^\dagger_{(i,j)}\sigma^\dagger_{(i,j+1)}\sigma_{(i-1,j)}\sigma_{(i-1,j+1)}
& \mbox{otherwise.}
\end{array}
\right. 
\end{equation}
Then the dual Hamiltonian, $H_{\sf G}=\Phi_\d(H_{\sf stripe})$, reads
\begin{equation}\label{gaugeh}
H_{\sf G}=-\Big[ \frac{\Delta}{2}\sum_{i+j={\sf even}} B_{{\sf d}\,(i,j)}
 +\frac{J}{2}\sum_{j=0}^{L_y-1}\sum_{i=0}^{L_x-2} 
\tau_{(i,j)}\Big]+{\rm h.c.}
\end{equation}
and it is unitarily equivalent \cite{cobanera2011} to the stripe model. In Appendix \ref{dual_vars} we write down explicitly the dual clock operators and show that it is possible to interpret the gauge theory Hamiltonian \eqref{gaugeh} as the Hamiltonian governing the collective modes of the stripe model.

\begin{figure}[t!]
\includegraphics[width=\columnwidth]{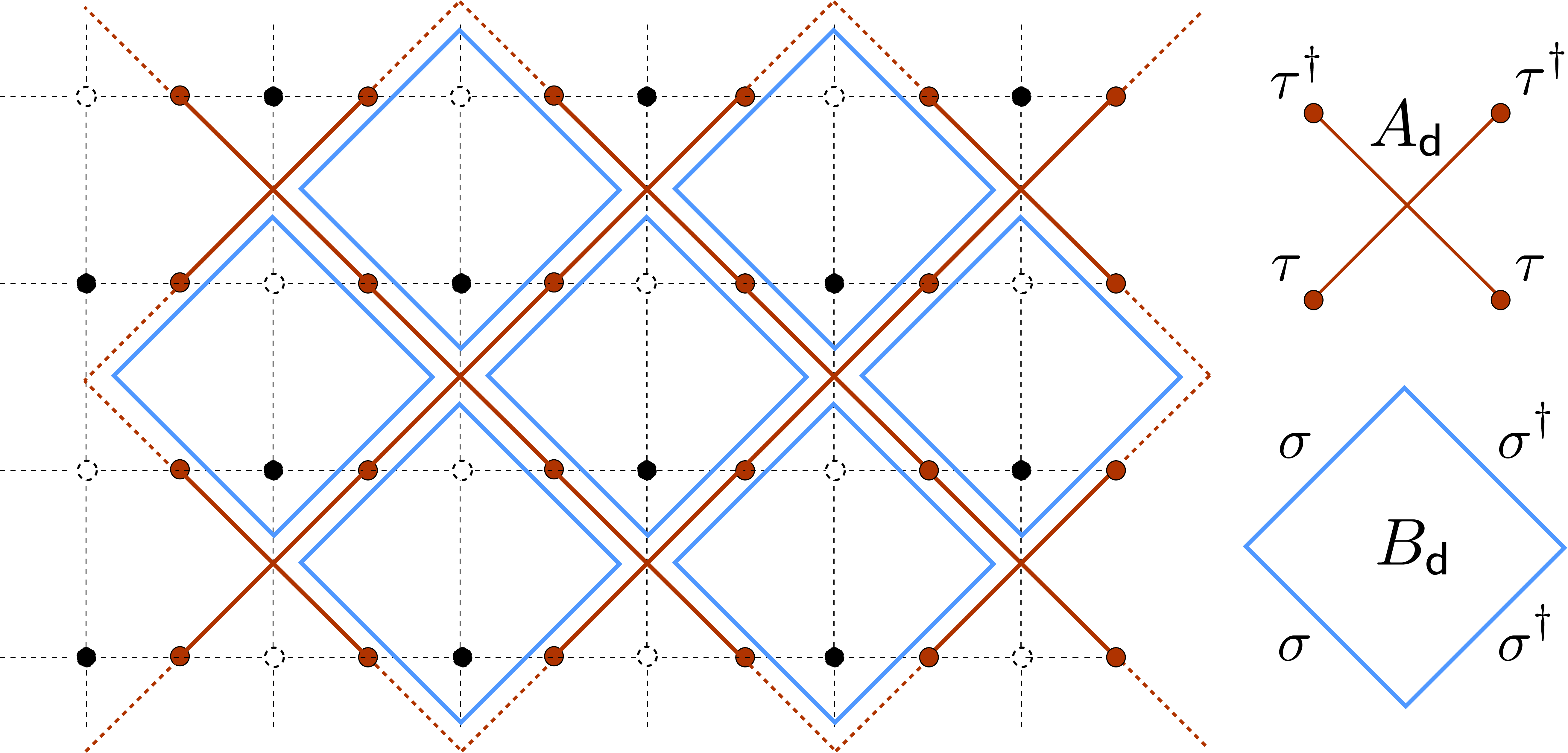}
\caption{The dual lattice on which the $Z_{2m}$ lattice gauge theory of Eq. \eqref{gaugeh} is defined. Clock operators now live on those links of the original square lattice which are marked by a blue dot. On this new lattice we find that in perturbation theory the physical interactions are given by the plaquette and star operators $B_{\sf d}$ and $A_{\sf d}$ defined in Eqs. \eqref{def_plaquette}, \eqref{Adual}. \label{fig_gauge_lattice}}
\end{figure}

Up to boundary terms, i.e. the incomplete plaquettes \(B_{(0,j)}\ (j=0,2,\dots,L_y-2)\),
and a redefinition of the lattice that places the clock degrees of freedom 
on links rather than sites, we recognize \(H_{\sf G}\) as the Hamiltonian of the \(\Z_{2m}\) lattice gauge theory 
studied in connection to the problem of confinement in QCD \cite{horn} (see Fig.~ \ref{fig_gauge_lattice}). 
The local symmetries $B_{(i,j)}$ of the stripe model, Eq. \eqref{upsilons},  
map under duality to
\begin{equation}\label{dualups}
B_{(i,j)}\dual A_{{\sf d}\,(i,j)},\qquad i+j={\sf even},
\end{equation} 
with
\begin{equation}\label{Adual}
A_{{\sf d}\,(i,j)}\equiv \tau_{(i,j)}\tau_{(i+1,j)}\tau^\dagger_{(i,j+1)}\tau^\dagger_{(i+1,j+1)}.
\end{equation}
As is guaranteed by the formalism, the unitary operators $A_{{\sf d}\,(i,j)}$,  $(i+j={\sf even})$, commute with the dual Hamiltonian $H_{\sf G}$. They correspond to the gauge symmetries of the $\Z_{2m}$ gauge theory and they have the interpretation of measuring the local density of external 
$\Z_{2m}$ charge. Hence our duality maps the magnetic fluxes in the stripe model,
as described by the Aharonov-Bohm operators 
$B_{(i,j)}$ in Eq. \eqref{upsilons}, to external 
$\Z_{2m}$ electric charges in the gauge theory.

At this point we can exploit Eq. \eqref{dualups} 
to compute the dual representation 
\(H^{{\sf pert}\, D}_{\sf stripe}=\Phi_\d(H^{\sf pert}_{\sf stripe})\)
of the perturbative Hamiltonian of Eq. \eqref{pert_stripe},
\begin{equation}
H^{{\sf pert}}_{\sf G}=- \sum_{i+j={\bf even}}\Big[ 
\frac{\Delta}{2}B_{{\sf d}\,(i,j)} + \frac{5J^4}{4G^3} A_{{\sf d}\,(i,j)}\Big]  
+ {\rm h. c.}\ .
\end{equation}
Remarkably, this is the Hamiltonian for the qudit toric code 
model\cite{bullock2007,schulz2012}. To clarify the notation,
notice that due to our definition of the plaquette operator Eq. \eqref{def_plaquette}, 
the two operators \(A_{{\sf d}\,(i,j)},B_{{\sf d}\,(i,j)}\) share one vertical link of the lattice, 
with \(A_{{\sf d}\,(i,j)}\) to the right and \(B_{{\sf d}\,(i,j)}\) to the left of that link (see Fig.~ \ref{fig_gauge_lattice}). This duality, however, is a non-local transformation with respect to the clock operators $\sigma$ and $\tau$. Thus, even if the spectrum of $H^{\sf pert}_{\sf stripe}$ is equivalent to the \(\Z_{\sf 2m}\) toric code, the stripe model in the Coulomb-dominated regime does not present topological order.

To assert that the phase diagram of the gauge theory and the stripe model are 
connected by the duality $\Phi_d$, we need to investigate
the effect of the duality mapping on the charge constraints of Eq. \eqref{stripes_const}.
Remarkably, the charge constraints are {\it holographic} \cite{holographic}, 
since they map to boundary constraints for \(H_{\sf G}\), 
\begin{equation}
S_jS_{j+1}\dual\sigma^\dagger_{(L_x-1,j)}\sigma^\dagger_{(L_x-1,j+1)},
\end{equation}
for \(\quad j=0,2,\cdots,L_y-2\). Then the physical states $\ket{\Psi}_{\sf d}=\Phi_{\sf d}\ket\Psi$ for  \(H_{\sf G}\),  seen as a 
dual representation of the stripe model, are characterized by the condition
\begin{equation}\label{holophysical}
\sigma^\dagger_{(L_x-1,j)}\sigma^\dagger_{(L_x-1,j+1)}\ket{\Psi}_{\sf d}=\ket{\Psi}_{\sf d}
\end{equation}
for \(\quad j=0,2,\cdots,L_y-2\), and not by the standard condition of gauge 
invariance [that is, invariance under the $A_{(i,j)}$,  $(i+j={\sf even})$].\cite{horn}
Despite this difference we will argue in the following that the stripe model and the $\mathbb{Z}_{2m}$ lattice gauge theory share the same phase diagram.

The dual gauge theory \(H_{\sf G}\) presents
unusual open boundary conditions. Since the charge constraints are holographic,
this is in perspective required to guarantee that the dual charge constraints
supported on the boundary commute with the dual Hamiltonian  \(H_{\sf G}\)
(just as the charge constraints commute with \(H_{\sf stripe}\)).

\begin{figure}[t!]
\includegraphics[width=\columnwidth]{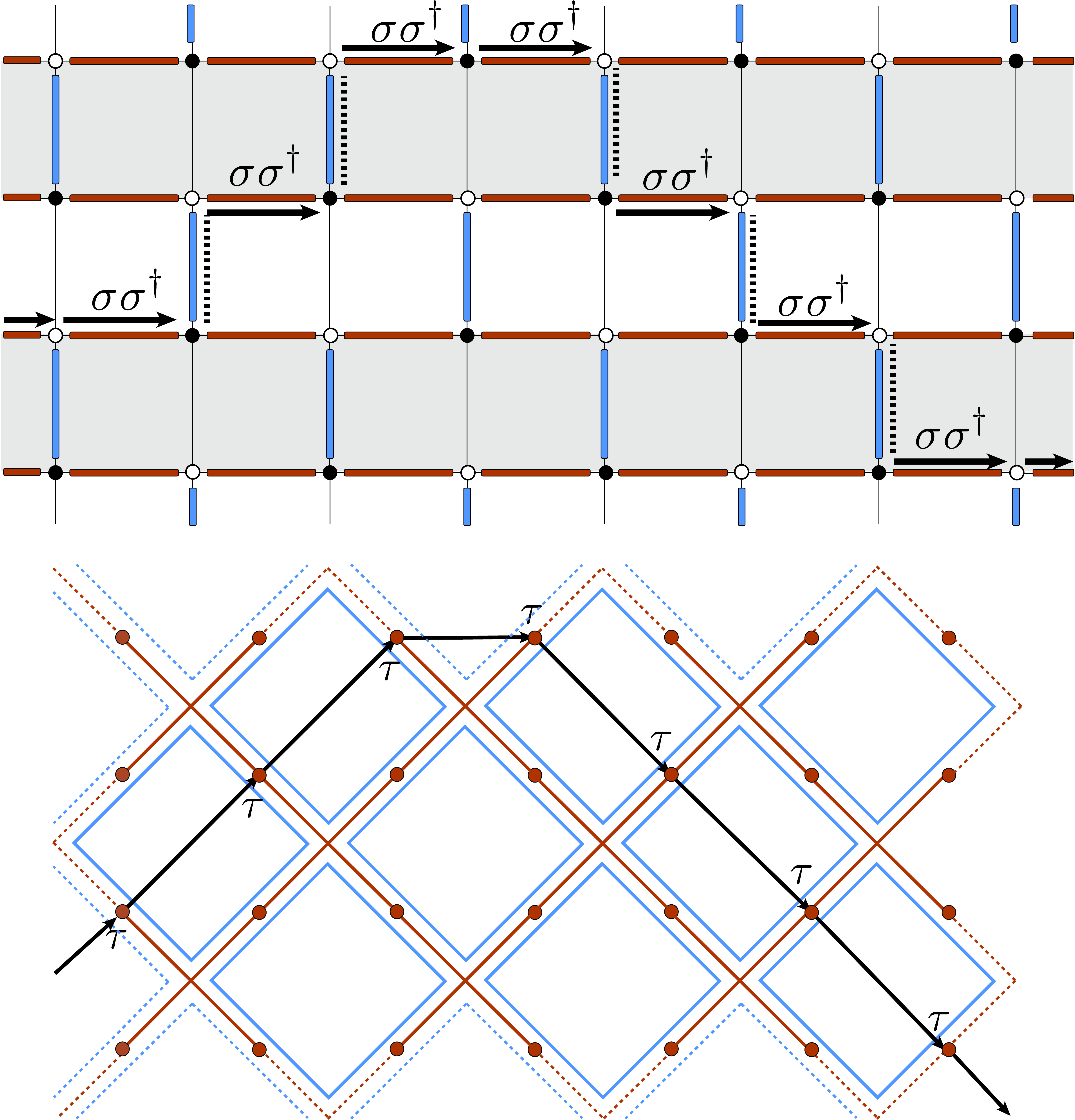}
\caption{The generalized order parameter (string tension) in both the original and dual lattices. \label{fig_string_tension}}
\end{figure}

However, the standard view that boundary conditions do not affect the phase diagram in the thermodynamic limit suggests in this case that the dual charge constraints do not affect the phase diagram of  \(H_{\sf G}\), which must then coincide with the standard phase diagram of the \(\Z_{2m}\) gauge theory. This view is strengthened
by the fact that the dual charge constraints commute with the gauge symmetries of 
\(H_{\sf G}\), and so the ground state of \(H_{\sf G}\) will
belong to the gauge-invariant sector even after the condition
Eq. \eqref{holophysical} is imposed. Finally this implies that the stripe model presents the same phase diagram independently on the choice of the charge of each FTI edge in the constraint \eqref{chargeconstraint}, thus in all the different physical sectors.

In view of these considerations, we can argue that the stripe
model shares the phase diagram of the \(\Z_{2m}\) gauge theory as described in Ref. \onlinecite{horn} (and references therein). It follows that there is indeed one second-order phase transition separating the Coulomb-dominated
from the Josephson-dominated regime. In the gauge-theory language this transition is understood as a 
confinement-deconfinement transition. In particular, the Coulomb-dominated regime of the stripe model is dual to to the deconfined phase of the gauge theory, while the Josephson-dominated regime corresponds
to the confined phase.

The phases of a gauge theory cannot be distinguished by a Landau order
parameter\cite{elitzur75}. However, for the \(\Z_{2m}\) gauge theory
dual to the stripe model, there exists a generalized order parameter,
the so-called {\it string tension}\cite{holographic}, which is non-zero in the confining phase 
and vanishes continuously, but non-analitically
at the transition point. The string tension is the expectation value of a string of $\tau$'s in the $\mathbb{Z}_{2m}$ gauge theory, which corresponds to an open string of tunneling operators $\sigma \sigma^\dag$ in the stripe model (see Fig.~ \ref{fig_string_tension}, analogously to the string operators creating charge excitations in the tile model. The ground-state expectation value of such string falls continuously, 
but non-analytically to zero at the second-order phase transition separating the 
Josephson-dominated regime (where it is different from zero) from the Coulomb-dominated 
regime. On the gauge theory side of the duality, the two phases can be distinguished also by a different scaling of the expectation value of the Wilson loops, which map to sets of $\tau$ operators in compact regions in the stripe model.

\section{Conclusions and Outlook}\label{sec_conclusion}

In summary, we have studied two-dimensional arrays of interacting parafermionic zero modes. Such exotic states form along the edge of fractional topological insulators, at the domain walls between proximity-induced superconducting and ferromagnetic pairing. The dynamic of these zero modes is dictated by two competing effects: the charging energy of each superconducting island and the fractional Josephson tunneling of quasiparticles between different islands.

The underlying fractional edge modes, which are originally described by a helical Luttinger liquid theory, determine crucially the possible lattice geometries and the physical constraints of these parafermionic systems. We have analyzed two possible architectures, the tile and the stripe model. They differ mainly for the fact that in the former the length of the edges is constant, while in the latter it scales with the total size of the architecture. We have discussed how this feature gives rise to different physics, despite the fact that the models are characterized by the same lattice of parafermions and the same local interactions. 

The difference is appreciated by exploiting a Jordan-Wigner transformation mapping the parafermionic operators into clock operators. Through this transformation the tile model is described by a Hamiltonian on a decorated square lattice whereas the stripe model becomes a compass model, with $\mathbb{Z}_{2m}$ symmetry, on a brick-wall lattice.

The tile model presents, at least in perturbation theory, the same topological order of the surface codes characterized by a $\mathbb{Z}_{2m}$ symmetry: it shows the same topological degeneracy of the ground state and the same anyonic excitations. Thus the system we described is a possible physical candidate to the realization of qudit surface code Hamiltonians. It is known that the ground state degeneracy of these systems suffers from thermal fragility\cite{dennis02,nussinov08}. However, we note that the intrinsic noise due to the presence of induced charges on the superconducting islands could help localize the anyonic excitations of the system, and thus to protect the information which may be encoded in the ground states.\cite{wootton11,stark11}

The stripe model provides instead a physical realization of the $\mathbb{Z}_{2m}$ lattice gauge theory, a toy model often exploited to study confinement-related problems in lattice field theory. The duality mapping between the stripe model and the lattice gauge theory is non-local in terms of the single clock degrees of freedom, but it is local in terms of the interactions. Unlike in the tile model, a toric code Hamiltonian can only be retrieved in the dual theory, where the operators are non-local. Interestingly enough the physical charge constraints of the FTI edges maps to an holographic constraint in the gauge theory which affects only boundary terms. 

To conclude, our work addressed the problem of finding topologically ordered phases in the phase diagram of these two-dimensional collections of topological defects. Comparing the results obtained for the two architectures, we can see that that it is not only the nature of the interactions between the defects (in this case, the $\Z_{2m}$ PF zero modes) that determines the presence of topological order, but also the intrinsic geometry of the topological phases originally generating the defects (in this case, the edges of the fractional topological insulators). Understanding the interplay between this two aspects is crucial to design topologically-ordered architectures.

\section*{Acknowledgements.}

We would like to thank C.W.J. Beenakker, D. Ferraro, N. Magnoli, G. Ortiz, R. Orus and B. Paredes for insightful discussions.
This project was supported by the Dutch Science Foundation NWO/FOM and an ERC Advanced Investigator Grant.

\appendix

\section{Description of the system through bosonization}\label{bosonization_appendix}

In this Appendix we summarize the main features of the bosonization description of our system and we provide an expression in terms of massless bosonic fields of the parafermion operators $\alpha$ and thus of the related interaction terms. We follow the approach in Refs. \onlinecite{stern2012} and \onlinecite{shtengel2012}, where more details can be found.

In absence of the interactions provided by the superconducting islands and the ferromagnetic insulators, the edge of the FTIs defining our systems, or, equivalently, the double edges of juxtaposed fractional quantum Hall layers with opposite polarization, can be described in terms of the Luttinger liquid Hamitonian proposed by Wen \cite{wen1990,lee1991}. In particular the massless edge modes are described by the following Hamiltonian:
\begin{equation}
 H_0 = \frac{mv}{2\pi}\int \de x \left[\left(\partial_x \varphi \right)^2 +\left(\partial_x\theta \right)^2  \right] ,
\end{equation}
where $v$ is the speed of the two counterpropagating modes and $\varphi,\theta$ are dual massless bosonic fields obeying the commutation relation:
\begin{equation}
 \left[\varphi\left(x_1,\mathsf{a},t \right),\theta\left(x_2,\mathsf{a'},t \right)  \right]= i \frac{\pi}{m} \delta_{\mathsf{a},\mathsf{a'}} \Theta\left( x_2-x_1\right),  \label{com}
\end{equation}
where $\Theta$ is the Heaviside step function. In particular for each FTI edge $\mathsf{a}$ it is possible to define two chiral bosonic fields
\begin{equation}
 \varphi_{L/R} \left(x\pm t,\mathsf{a}\right) \equiv \varphi\left(x,t,\mathsf{a}\right)  \mp \theta\left(x,t,\mathsf{a}\right) ,
\end{equation}
in such a way that the left and right fermionic modes, with opposite spin polarization, are defined by the operators $\psi_{L/R}(x,t,\mathsf{a})=\eta_\mathsf{a} \e^{im\varphi_{L/R}(x\pm vt,\mathsf{a})}$ where $\eta_\mathsf{a}$ are fermionic Klein factors.
The charge density associated with each edge is $\rho=\partial_x \theta / \pi$, thus, in a closed edge with length $\mathcal{L}$, the total charge of the edge is related to the boundary conditions of the $\theta$ field:
\begin{equation} \label{boundtheta}
 \pi q_{\sf tot}(\mathsf{a}) = \theta\left(x + \mathcal{L},\mathsf{a}\right) - \theta\left(x,\mathsf{a} \right) 
\end{equation}
and analogous conditions relate the field $\varphi$ with the spin density \cite{stern2012}.

For each edge the interaction terms corresponding to the proximity induced superconducting coupling and the backscattering give rise to the interaction Hamiltonian:
\begin{equation}
 H_I \propto \int \de x \left[ -g_S(x) \cos\left( 2m\varphi\right) - g_F(x) \cos\left(2m\theta \right)  \right] 
\end{equation}
where $g_S$ and $g_F$ describe respectively the position dependence of the induced superconducting and ferromagnetic couplings.

By selecting a position in the bulk of either a superconducting or a ferromagnetic segment of the edge, if the couplings $g$ are strong enough, one can consider respectively the fields $\varphi$ and $\theta$ as pinned to the semiclassical minima $\varphi_k,\theta_k=0, \pi/m, 2\pi/m, \ldots, (2m-1)\pi/m$. Adopting this approximation and considering the limit of sharp transitions between superconducting and ferromagnetic regions, the parafermion operators can be written as:
\begin{eqnarray}
\alpha_{2k-1,\mathsf{a}} = \kappa_\mathsf{a} e^{i\left(\varphi_{k,\mathsf{a}} - \theta_{k,\mathsf{a}} \right) } \\
\alpha_{2k,\mathsf{a}} = \kappa_\mathsf{a} e^{i\left(\varphi_{k+1,\mathsf{a}} - \theta_{k,\mathsf{a}} \right) } 
\end{eqnarray}
where $k=1,\ldots, M$ labels the ferromagnets and the superconductors along the edge $\mathsf{a}$ and the tile and stripe models are characterized respectively by $M=4$ and $M=L_x$. The fractional Klein factors $\kappa_{\mathsf{a}}$ enforce the correct commutation rules \eqref{commrules} and \eqref{commrules2}. This definition of the parafermionic modes is not unique (see Refs. \onlinecite{stern2012} and \onlinecite{shtengel2012} for more detail) but it suffices to our purposes. 
Finally, for a complete description of the system, it is necessary to take into account the correct boundary conditions. 

Through this definition of the parafermionic operators it is easy to derive Eqs. (\ref{prop1},\ref{prop2},\ref{commrules},\ref{commrules2}) and verify that the tunneling operators assume the form $\e^{-i\left( \varphi_{k+1,\mathsf{a}} - \varphi_{k,\mathsf{a}} \right) }=P_{\{{\sf a},2k-1\},\{{\sf a},2k\}}$; thus we recover the usual form for the fractional Josephson interaction \eqref{JJham}:
\begin{equation}
  H_J = -J \cos \left(\varphi_{k+1,\mathsf{a}} - \varphi_{k,\mathsf{a}} -\frac{\delta}{2m} \right) 
\end{equation}
Moreover the tunneling string operator $\Sigma_\mathsf{a}$ defined in \eqref{stringoperators} for the two models becomes $\Sigma_\mathsf{a}=\exp\left( -{i}\left(\varphi_{M+1,\mathsf{a}} - \varphi_{1,\mathsf{a}} \right) \right)$, emphasizing the relation between the boundary conditions of the field $\varphi$ and the magnetic flux enclosed by the FTI edges.
The boundary condition \eqref{alpha_boundary} assumes a natural form in the bosonized description due to the boundary relation \eqref{boundtheta} since:
\begin{equation} \label{ultimateboundary}
\alpha_{\left\lbrace \mathsf{a},2M+1\right\rbrace } = \kappa_\mathsf{a} \e^{i\left(\varphi_{M+1,\mathsf{a}} - \theta_{M+1,\mathsf{a}} \right) } =  \e^{-i\pi q_{\sf a}} \alpha_{\left\lbrace \mathsf{a},1\right\rbrace } \Sigma^\dag_\mathsf{a}.
\end{equation}
Once we apply the parafermionic Jordan-Wigner transformation (\ref{JW1},\ref{JW2}) to map the system in a quantum clock model, the previous boundary conditions are translated in the following relations:
\begin{align}
 & e^{i\pi q_\mathsf{a}}=e^{i\left( \theta_{M+1,\mathsf{a}} - \theta_{1,\mathsf{a}}  \right) }=\prod\limits_{k=1}^{M} \tau^{\dag}_{k+1} \\
 & \Sigma_\mathsf{a}=e^{-{i}\left(\varphi_{M+1,\mathsf{a}} - \varphi_{1,\mathsf{a}} \right) }=\sigma^\dag_{M+1}\sigma_1
\end{align}
which generalize the boundary conditions (\ref{sigmaconst},\ref{tauconst},\ref{tauconst2}).

\section{Collective modes of the stripe model and the \(Z_{\sf 2m}\) gauge
theory}\label{dual_vars}

It is interesting to reinterpret the duality for the stripe model
in terms of collective modes. Let us define a new set of clock degrees of 
freedom as 
\begin{equation}
\hat{\sigma}_{(i,j)}\equiv \Phi_\d^{-1}(\sigma_{(i,j)}), \quad 
\hat{\tau}_{(i,j)}\equiv \Phi_\d^{-1}(\sigma_{(i,j)}).
\end{equation}
Here \(\Phi_\d^{-1}\) is the duality mapping the \(Z_{\sf 2m}\) gauge theory to 
the stripe model, obtained from Eqs. \eqref{d1} and \eqref{d2} by reading all arrows in 
reverse. As we will show shortly, the dual variables 
\(\hat{\sigma}_{(i,j)},\hat{\tau}_{(i,j)}\) 
are non-local operators when written in terms of the clock degrees of freedom 
\(\sigma_{(i,j)}, \tau_{(i,j)}\). 
The duality mapping \(\Phi_\d^{-1}\) shows that these collective modes of 
the stripe model are governed by the \(Z_{\sf 2m}\) gauge theory Hamiltonian, since
\begin{eqnarray}
&&H_{\sf stripe}=\Phi_\d^{-1}(H_{\sf G})=\\
&-&\Big[ \frac{\Delta}{2}\sum_{i+j={\sf even}} \widehat{B}_{{\sf d}\,(i,j)}
+\frac{J}{2}\sum_{j=0}^{L_y-1}\sum_{i=0}^{L_x-2} 
\hat{\tau}_{(i,j)}\Big]+{\rm h.c.}\ ,\nonumber
\end{eqnarray}
with \(\widehat{B}_{{\sf d}\,(i,j)}\) defined just as in Eq. \eqref{def_plaquette} 
up to the substitution \(\sigma_{(i,j)}\rightarrow \hat{\sigma}_{(i,j)}\). 
It follows that the stripe model realizes the \(\Z_{2m}\) gauge theory in terms
of its collective modes \(\hat{\sigma}_{(i,j)},\hat{\tau}_{(i,j)}\).

To compute the dual variables explicitly it is necessary to extend the 
bond algebra of the \(\Z_{2m}\) gauge theory by adding the boundary operators
\(\tau_{(L_x-1,j)}\ (j=0,\dots, L_y-1)\), 
\(\sigma^\dagger_{(0,0)}\), and  \(\sigma^\dagger_{(i,0)}\sigma_{(i-1,0)}\ (i=1,\dots, L_x-1)\)
to its list of bond generators. We also need to determine an algebraic extension of the 
duality mapping to these extra bonds,
\begin{eqnarray}
\sigma^\dagger_{(0,0)}&\idual&\tau_{(0,0)},\\
\sigma^\dagger_{(i,0)}\sigma_{(i-1,0)}&\idual& \tau_{(i,0)},\\
\tau_{(L_x-1,j)}&\idual& \sigma_{(L_x-1,j)}.
\end{eqnarray}
This completes the preliminaries. It follows that
\begin{equation}
\hat{\sigma}_{(i,j)}=\tau_{(0,j)}^\dagger\tau_{(1,j)}^\dagger\dots \tau_{(i,j)}^\dagger,
\end{equation}
and 
\begin{equation}
\hat{\tau}_{(i,j)}=\left\{
\begin{array}{ll}
\sigma_{(L_x-1,j)}& \mbox{if}\ i=L_x-1,\\
\sigma_{(i,j)}\sigma_{(i+1,j)}^\dagger
& \mbox{otherwise.}
\end{array}
\right. 
\end{equation}
It is possible to check directly that the dual variables satisfy
the correct algebra for clock degrees of freedom.  


\begin{thebibliography}{99}
\bibitem{LxLy} Note that the geometry of the array constrains $L_y$ to be even and $L_x$ to be a multiple of four.
\end{thebibliography}
\end{document}